\RequirePackage{rotating}
\documentclass[a4paper,fleqn,usenatbib,useAMS]{mnras}
\usepackage{inputenc}
\usepackage{url}
\usepackage{hyperref}
\usepackage{graphicx}
\usepackage{amsmath}
\usepackage{amssymb}
\usepackage[labelfont=bf,labelsep=period]{caption}
\usepackage{subcaption}
\usepackage{rotating}
\usepackage{float}
\usepackage{multirow}
\usepackage{xcolor}
\usepackage{footmisc}
\usepackage{array,ragged2e}
\usepackage{hhline}
\usepackage{threeparttable,tablefootnote}
\usepackage{threeparttablex}
\usepackage{lscape}

\def\xmm{{\it XMM-Newton}}

\def\chandra{{\it Chandra}}
\def\swift{{\it Swift}}
\def\swiftng{{\it Neil Gehrels Swift Observatory}}

\def\epicmos1{{EPIC-MOS1}}
\def\epicmos2{{EPIC-MOS2}}
\def\epicmos{{EPIC-MOS}}

\def\athena{\textit{Athena}}
\def\erosita{\textit{eROSITA}}

\def\webpimms{{\small WEBPIMMS}}
\def\ciao{{\small CIAO}}

\def\apj{ApJ}
\def\mnras{MNRAS}
\def\nat{Nat}

\def\araa{ARA\&A}                % "Ann. Rev. Astron. Astrophys."
\def\aap{A\&A}                   % "Astron. Astrophys."
\def\aj{AJ}                      % "Astron. J."
\def\apjs{ApJS}                  % "Astrophys. J. Suppl. Ser."
\def\pasp{PASP}                  % "Publ. Astron. Soc. Pac."
\def\apjl{ApJ}                   % letter at ApJ

\def\eg{\textit{e.g.}}

\def\msun{\hbox{$M_{\odot}$}}
\def\ergps{erg~s$^{-1}$}

\def\n5907{NGC\,5907 ULX1}
\def\n300{NGC\,300 ULX1}

\title[The Hunt for PULXs]{The Hunt for Pulsating Ultraluminous X-ray Sources}

\author[X. Song et al.]{\parbox{7.in}{
X. Song,$^{1,2}$\thanks{E-mail: xsong@pulsarastronomy.net (XS)}
D. J. Walton,$^{1}$
G. B. Lansbury,$^{1}$
P. A. Evans,$^{3}$
A. C. Fabian,$^{1}$
H. Earnshaw,$^{4}$
T. P. Roberts$^{5}$
}
\\
\\
$^{1}$Institute of Astronomy, University of Cambridge, Madingley Road, Cambridge, CB3 0HA, UK\\
$^{2}$Jodrell Bank Centre for Astrophysics,
Department of Physics and Astronomy, University of Manchester, M13 9PL, UK\\
$^{3}$Department of Physics and Astronomy, University of Leicester, Leicester, LE1 7RH, UK\\
$^{4}$Cahill Center for Astronomy and Astrophysics, California Institute of Technology, Pasadena, CA 91125, USA\\
$^{5}$Centre for Extragalactic Astronomy, Durham University, Department of Physics, South Road, Durham DH1 3LE, UK
}

\date{Accepted 2019 October 24. Received 2019 October 20; in original form 2019 September 2.}

\pubyear{2019}

\begin{document}
\label{firstpage}
\pagerange{\pageref{firstpage}--\pageref{lastpage}}
\maketitle

\begin{abstract}
Motivated by the recent discoveries that six Ultraluminous X-ray Sources (ULXs) are powered by highly super-Eddington X-ray pulsars, we searched for additional pulsating ULX (PULX) candidates by identifying sources that exhibit long-term flux variability of at least an order of magnitude (a common feature seen in the 6 known PULXs, which may potentially be related to transitions to the propeller regime). Expanding on previous studies, we used the available fluxes from \xmm, \swift\ and \chandra, along with carefully computed upper limits in cases of a non-detection, to construct long-term lightcurves for a sample of 296 ULXs selected from the \xmm\ archive. Among these 296, we find 25 sources showing flux variability larger than a factor of 10, of which 17 show some evidence for (or are at least consistent with) exhibiting bi-modal flux distributions, as would be expected for sources undergoing propeller transitions. These sources are excellent candidates for continued monitoring programs to further test for this behaviour. There are 3 sources in our final sample with fluxes similar to NGC 5907 ULX1, currently the faintest known PULX, which would also be good targets for deeper observations with current facilities to search for pulsations. For the rest of the PULX candidates identified here, the next generation of X-ray telescopes (such as \athena) may be required to determine their nature owing to their lower peak fluxes.
\end{abstract}


\section{Introduction}

Ultraluminous X-ray sources (ULXs) are off-nuclear extragalactic objects with X-ray luminosities higher than $10^{39}$\,\ergps, roughly the Eddington luminosity ($L_{\rm{Edd}}$) for a standard stellar remnant black hole (BH; $\sim$10\,\msun) \citep{Kaaret2017}. Although it was long believed that ULXs were mostly BHs, coherent pulsations have been recently found from 6 ULXs: M82 X-2 \citep{Bachetti2014}, NGC\,7793 P13 \citep{Furst2016, Israel2017}, NGC\,5907 ULX1 \citep{IsraelBelfiore2017}, NGC\,300 ULX1 \citep{Carpano2018}, NGC\,1313 X-2 \citep{Sathyaprakash19} and M51 ULX7 \citep{Rodriguez19}, indicating that some ULXs are neutron stars (NSs). These pulsating ULXs (PULXs) show extreme observational characteristics. Among them, NGC\,5907 ULX1 is the most luminous NS found so far with a luminosity of about 10$^{41}$\,\ergps, $\sim$500 times higher than the corresponding Eddington limit of NSs \citep{IsraelBelfiore2017,Furst2017}. Furthermore, monitoring of the pulse period shows that they are all spinning up. For NGC\,300 ULX1, the most extreme case, the spin period changed from 32 seconds to about 19 seconds in 2 years from 2016 to 2018 \citep{Carpano2018, Bachetti2018}. Another ULX, M51 ULX8, has also been identified as a likely neutron star accretor through the detection of a potential cyclotron resonant scattering feature \citep[CRSF;][]{Brightman2018}, although pulsations have not yet been detected from this source.

Currently there is significant debate over how these neutron star ULXs are able to reach such extreme luminosities. Although there must be some degree of anisotropy to the radiation field to see pulsations \citep[\eg][]{Basko1976}, these systems do not appear to be strongly beamed, as their pulse profiles are all nearly sinusoidal \citep{Bachetti2014, Furst2016, Israel2017, IsraelBelfiore2017, Carpano2018}. The debate primarily focuses on the magnetic fields of these systems. One possibility invokes strong, magnetar-level magnetic fields ($B \sim 10^{14}$\,G; \eg\ \citealt{Eksi2015, DallOsso2015, Mushtukov2015}). Such extreme fields reduce the electron scattering cross section \citep{Herold1979}, and in turn increase the effective Eddington luminosity. If dipolar, fields of this strength would truncate the accretion flow at large radii, although the presence of higher-order (\eg\ quadrupolar) components to the field close to the surface of the neutron star may ease this constraint to some extent (\eg\ \citealt{IsraelBelfiore2017}). However, other authors have instead argued for low magnetic fields (potentially as low as $B \sim 10^9$\,G) based on the ratio of the spin-up rate to the luminosity, which is an order of magnitude lower than typical X-ray pulsars and may imply that the disk extends close to the neutron star surface (\eg\ \citealt{Kluzniak2015}). If this is the case, the extreme luminosities would need to be produced by a highly super-Eddington accretion disk that extends close to the accretor, similar to super-Eddington accretion onto a black hole (\citealt{King2016}). The two potential direct constraints on PULX magnetic fields from CRSFs paint a mixed picture; \cite{Brightman2018} identify the feature in M51 ULX8 as a proton CRSF, implying a magnetar-level field of $B \sim 10^{15}$\,G (although see also \citealt{Middleton19}), while \citet{WaltonS} present a potential electron CRSF in NGC\,300 ULX1, implying a much more moderate field of $B \sim 10^{12}$\,G (however, see also \citealt{Koliopanos2019}).

With so few examples currently known, identifying additional pulsar/neutron star ULXs will necessarily play a major role in furthering our understanding of these remarkable systems. However, this is complicated by the fact that in half of the known PULXs the pulsations are observed to be transient. Furthermore, the pulsations can be challenging to detect when present owing to the combination of the low count-rates from ULXs and the fact that significant period derivatives are seen in the known PULXs. We therefore require additional means to identify promising PULX candidates.

In addition to hosting neutron star accretors, a number of the known PULXs also have other characteristics in common, particularly in terms of their long-term variability properties. M82\,X-2, NGC\,7793 P13 and NGC\,5907 ULX1 all exhibit unusual `off' states in which their fluxes drop by factors of $\sim$100 (or more) relative to their typical ULX states \citep{Motch2014, Walton2015, Brightman2016}, potentially offering a means to identify new PULX candidates (NGC\,300 ULX1 is also known to exhibit high-amplitude variability, but high-cadence monitoring has only begun relatively recently for this source). \cite{Tsygankov2016} proposed that these off-states are related to the propeller effect, in which the magnetic field suddenly acts as a barrier to accretion and shuts off the observed luminosity. Based on this possibility, \cite{Earnshaw2018} searched the \xmm\ archive for other sources that exhibit long-term variability in excess of an order of magnitude in flux, and found another highly variable ULX that shows evidence for a bi-modal flux distribution, as would be expected for a neutron star transitioning into and out of the propeller regime. Recently, \citet{Brightman2019} found that the off-states in M82\,X-2 appear to be associated with its super-orbital period of $\sim$60 days, providing another potential explanation to the off-states observed in some cases.

In this work, we expand on the preliminary analysis presented in \cite{Earnshaw2018} and compile data from all of the major soft X-ray observatories -- \xmm\ (\citealt{XMM}), \chandra\ (\citealt{CHANDRA}) and the \swiftng\ (hereafter \swift; \citealt{SWIFT}) -- for a large sample of ULXs to facilitate an expanded search for highly variable sources and identify further PULX candidates. The structure of the paper is as follows: Section\,\ref{sec:method} presents the data assembly for the ULX samples from the three telescopes. In Section\,\ref{sec:analysis}, we explain the refinements on the fluxes and upper limits to select highly variable samples. We discuss these selected ULXs based on their lightcurves in Section\,\ref{sec:discu}. The conclusion follows in Section\,\ref{sec:con}.

\section{Data Assembly}
\label{sec:method}

\subsection{The ULX Sample}
We began with the latest available ULX catalogue, presented in \cite{Earnshaw19}, an update of the ULX catalogue compiled by \cite{Walton2011}. This is based on observations with the \xmm\ observatory, and was compiled by cross-correlating the fourth data release of the 3XMM Serendipitous Survey (DR4; \citealt{Rosen2016})\footnote{\url{https://xmmssc-www.star.le.ac.uk/Catalogue/3XMM-DR4/}} with the Third Reference Catalogue of Bright Galaxies (RC3, \citealp{Vaucouleurs1991}) and the Catalogue of Neighbouring Galaxies (CNG, \citealp{Karachentsev2004}). The \cite{Earnshaw19} catalogue includes 340 ULX candidates, considering only sources that have luminosities higher than $10^{39}$\,\ergps. We note that of the six known PULXs, only NGC\,5907 ULX1 and M51 ULX7 are present in this catalogue; M82 X-2 is blended with X-1 for \xmm, while the earliest \xmm\ observation of NGC\,7793 P13 only became publicly available in 2013 and NGC 300 ULX1 only reached ULX luminosities in 2016, both of which are after the cutoff for 3XMM-DR4. Finally, NGC\,1313 X-2 is formally located outside of the D25 isophote listed for NGC\,1313 in the RC3 catalogue, which is used to mark the extent of the galaxy in \cite{Earnshaw19}. Of the 340 ULX candidates, 296 sources have at least two observations in the combined \xmm, \swift\ and \chandra\ archives, allowing for at least a crude assessment of the level of variability observed. \citet{Earnshaw19} note that about $\sim$24\% of their sample of ULX candidates are estimated to be unidentified non-ULX contaminants (primarily background quasars). However, our focus is on highly variable sources, so it is worth noting that our source selection procedure is likely biased \textit{against} such sources, as background quasars do not typically exhibit the level of variability we are interested in on the timescales typically covered by the available lightcurves (e.g. \citealt{Paolillo2017}). In the following sections, we outline our data assembly procedure for building up long-term lightcurves of these sources. In general, throughout this work we refer to individual sources with their 3XMM-DR4 source identifications (SRCID), but where relevant we also give the full source name.

\subsection{\textit{XMM-Newton} Data}
\label{sec:xmm}

For each source considered, we updated the \xmm\ data to incorporate all of the observations included in 3XMM-DR7 (Data Release 7, data publicly available before 31 December 2016)\footnote{\url{http://xmmssc.irap.omp.eu/Catalogue/3XMM-DR7/3XMM_DR7.html}} to construct the long-term lightcurves. 
 
The 3XMM-DR7 catalogue provides \xmm\ fluxes in the 0.2--12\,keV band, based on data from the EPIC detectors (pn, MOS1 and MOS2; \citealt{XMM_PN, XMM_MOS}). However, the 3XMM catalogue does not provide information for cases in which a known source was observed but not detected, and such non-detections are of significant interest for our work. When this occurred, we therefore computed upper limits on the flux at the source position, as described in Section \ref{sec:upl}.

\subsubsection{Upper Limit Determination}
\label{sec:upl}

Due to the large amount of data we were dealing with, we initially used the FLIX tool\footnote{\url{http://www.ledas.ac.uk/flix/flix3}} to compute approximate 3$\sigma$ upper limits prior to 3XMM-DR5 (data publicly available before 31 December 2013; the current FLIX archive does not yet include data for later 3XMM releases). FLIX computes upper limits based on the exposure and the overall background level of an observation following the calculations outlined in \cite{Carrera2007}. However, while convenient, the FLIX upper limits can be significantly underestimated, so we only used these as an initial step to select highly variable candidates; the calculations performed by FLIX do not allow for the possibility of weak (but not significantly detected) source emission, nor potential local contamination from, \eg\ diffuse emission in the host galaxy or the PSF wings of bright nearby sources (\eg\ the central AGN) which results in an underestimation of the background flux from which the upper limit is derived.

We also therefore computed our own 3$\sigma$ upper limits for observations not covered in FLIX, as well as any sources selected in our initial search for strong variability based on the FLIX upper limits (see below). These manual upper limits were calculated by performing aperture photometry based on the method in \citet{Kraft1991} after carefully selecting source and background regions, as these can be critical for the determination of the upper limits. The source region used was a circle of radius 10, 15 or 20$''$, chosen on a case-by-case basis as a balance between avoiding nearby source contamination and including a reasonable fraction of the source emission. The background regions were selected to mimic the environment in which the source resides. For isolated sources, the background region was chosen to avoid other sources or background emission, and for sources close to another bright source, i.e.\ within its point spread function (PSF), the background region was selected to be at the same radial distance from the bright source.

These calculations gave the upper limits in raw counts. To determine the corresponding count-rate, the upper limit on the raw counts was divided by the exposure time at the source position, taken from the exposure map for that particular observation (which accounts for vignetting). We also corrected the upper limits for the fraction of the PSF outside of the source region (with the exact correction depending on the region size used), based on the fractional encircled energy at 1.5 keV, as this is where the effective area curves peak for the individual EPIC detectors\footnote{\label{ref:psf}\url{https://xmm-tools.cosmos.esa.int/external/xmm_user_support/documentation/uhb/onaxisxraypsf.html}}. The count-rates were then in turn converted to flux using the \webpimms\ tool\footnote{\label{ref:web}\url{https://heasarc.gsfc.nasa.gov/cgi-bin/Tools/w3pimms/w3pimms.pl}} (\citealt{PIMMS}), which accounts for the effective area and responses of the telescope and detectors. We assumed a generic spectral shape for this conversion, with a power-law photon index ($\Gamma$) of 1.7 and an absorption column density ($N_H$) of 3$\times 10^{20}$ cm$^{-2}$, which is consistent with that used by 3XMM-DR7 to compute source fluxes \citep{Rosen2016}. We noted that ULXs typically have softer spectra \citep[see \eg][]{Gladstone2009} but the spectral shape assumed should not affect the relative fluxes we are interested in. These calculations were performed for each of the EPIC detectors in turn, and we selected the minimum upper limit among the three to give the tightest constraint. Given the more detailed treatment of the individual sources, we consider these upper limits to be more robust than those returned by FLIX.

\subsection{\textit{Swift} Data}

We also compiled data from the XRT (\citealt{Burrows2005}) on board \swift. Although \swift\ typically monitors sources with larger numbers of short snapshot observations (typically $\sim$2\,ks exposure), such that the individual observations do not have the same sensitivity as those taken with \xmm\ and \chandra, the substantial temporal coverage provided by these data are of great benefit for a number of the sources considered.

We extracted the \swift\ data for our ULX sample using the standard lightcurve pipeline \citep{Evans2009}. This provides either XRT count-rates (if a source is detected) or 3$\sigma$ upper limits on the count-rate (if it is not) in the 0.3--10.0\,keV band as a function of time. The latter are calculated following the method described in \cite{Kraft1991}, similar to our manually calculated \xmm\ upper limits. We adopted a 5-day binning method to the fact that ULXs are extragalactic, and therefore typically quite faint. It is possible that some ULXs may vary significantly on shorter timescales than this (e.g. \citealt{Walton2015}), however in reality the chance of having multiple \swift\ observations within 5 days is rather low. With its default settings, the pipeline performs centroiding and applies a dynamic source region to maximise the signal-to-noise ratio (SNR) at a given position (with the appropriate PSF corrections applied). In most cases this is desirable, and so we kept these settings. However, for our sources of interest, we found that in a relatively small number of cases this resulted in misidentification and/or contamination from other nearby bright sources (see Section \ref{sec:analysis}). For these cases, we re-ran the \swift\ pipeline with centroiding turned off and a manually specified maximum source region size to address these issues. To construct the final lightcurve, the XRT count-rates/limits were converted to fluxes using \webpimms\footref{ref:web}. We again used $N_{\rm{H}}$ of 3$\times 10^{20}$~cm$^{-2}$ and $\Gamma$ of 1.7 for the conversion factor, the same as applied in 3XMM-DR7 \citep{Rosen2016}.

\subsection{\textit{Chandra} Data}

\chandra\ has the best imaging resolution (better than 1$''$ on-axis) of any X-ray mission flown to date, making it very efficient at detecting faint point sources. Observations with \chandra\ are therefore particularly useful for identifying blended sources that might not be resolved by \xmm\ and \swift, and potentially for constraining low-flux states. We compiled the available data from the latest Chandra Source Catalogue Release (version 2.0, hereafter CSC2), which contains publicly available data prior to 2014 observed with either the ACIS (\citealt{CHANDRA_ACIS}) or HRC (\citealt{CHANDRA_HRC}) detectors. We note that, at the time of writing, the current release contains $\sim$90\% of the full dataset to be included in CSC2, but only a few specific fields are still missing which are not generally of relevance here.

\begin{figure}
\centering
\includegraphics[width=\linewidth]{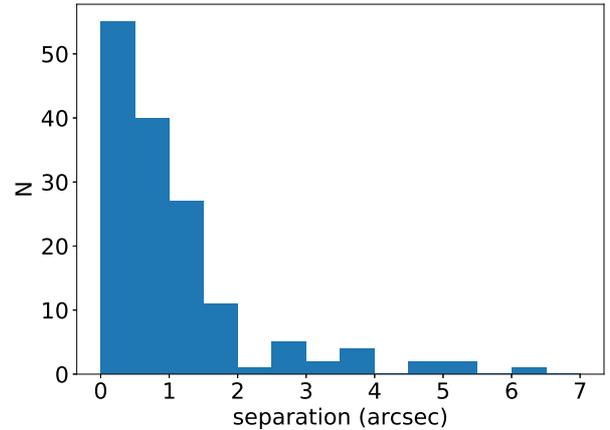} 
\caption{The distribution of the separation for the matched sources between the source positions of \xmm\ and \chandra. The bin size is set to be 0.5$''$.}
\label{fig:cxo_sep}
\end{figure}

To extract the \chandra\ data, we searched for sources within 10$''$ of the \xmm\ source position in the available CSC2 data. The distribution of the separation between the XMM and the Chandra positions of matched sources is shown in Figure \ref{fig:cxo_sep}. The separation between the \xmm\ and \chandra\ position peaks within a few arcseconds, which shows that the \chandra\ sources are very likely to be the true counterparts of the \xmm\ ones. For each detected source, CSC2 provides fluxes for individual observation (under the `Per Observation' tab in the catalogue). It includes a variety of model-dependent source fluxes in the 0.5--7\,keV band for ACIS observations (0.1--10\,keV for HRC). Where available, we used the fluxes calculated using an absorbed power-law model, which assumes $\Gamma$ = 2 and a (position-dependent) Galactic absorption column from the NRAO survey (given by the CIAO command \texttt{prop\_golden}\footnote{\url{http://cxc.harvard.edu/ciao/ahelp/colden.html}}). Although this is not the same as the model assumed when calculating the \xmm\ and \swift\ fluxes, we did not attempt to correct the \chandra\ fluxes owing to the strong time-dependence of the \chandra\ instrumental responses (related to the build-up of the well-known contaminant on the ACIS detectors; \citealt{Plucinsky18}). Tests with the latest \chandra\ responses suggest that, for a given count-rate, the different models should only result in differences of $\lesssim$20\% in the fluxes inferred, a small effect given the level of variability we are searching for (see Section \ref{sec:analysis}). 

Similar to our approach with \xmm, we also manually computed upper limits for any sources that were observed but not detected (determined using \texttt{find\_chandra\_obsid}). Note that, owing to \chandra's low background and superior sensitivity to faint point sources, this was a much rarer occurrence than with \xmm. For these calculations, we followed the same basic approach as outlined in Section \ref{sec:upl}). Source and background counts were extracted using the \texttt{srcflux} command in \ciao\ (which accounts for the time-dependent nature of the \chandra\ responses) with manually defined source and background regions, and converted to count-rates by using the observation exposure. We chose a fixed source region size of 3$''$, which includes more than 99\% of the \chandra\ PSF\footnote{\url{http://cxc.harvard.edu/proposer/POG/html/chap4.html}}.
The factor to convert count-rates to fluxes was given by \texttt{srcflux}. In these cases, we did assume a spectral model consistent with that used for the \xmm\ and \swift\ data (i.e. $\Gamma$ = 1.7 and $N_{\rm{H}}$ = 3$\times 10^{20}$ cm$^{-2}$).
Finally, we also noted that there were 8 entries in CSC2 for sources considered here that had measured count rates, but for which the flux conversion had not been applied. In these cases we converted the count rates into fluxes ourselves, again using conversion factors calculated with the \texttt{srcflux} command assuming the above model (for consistency with the \xmm\ and \swift\ data). 

\section{Selecting Highly Variable ULXs}
\label{sec:analysis}

Having assembled a large quantity of data on our ULX candidates, we performed a series of sanity checks in order to refine the sample and ensure that we were selecting ULXs that genuinely show high levels of variability, as outlined below.

\subsection{\textit{Chandra} Imaging}

\begin{figure}
\centering
\includegraphics[width=\linewidth]{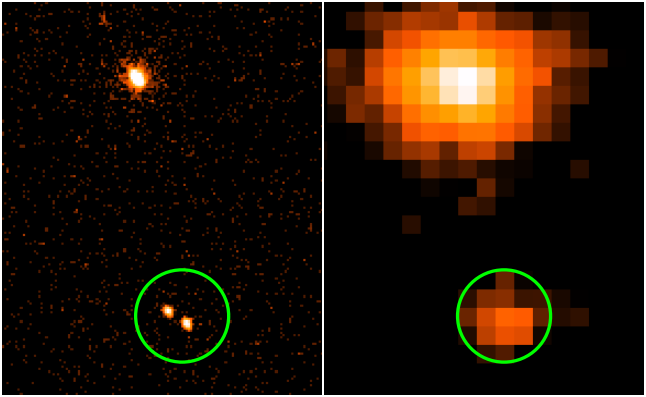} 
\caption{\chandra\ (left, obsID 2950) and \xmm\ PN image (right, obsID 0150280501) of SRC~901. The green 10$''$ circle is centred at the \xmm\ source position, which encircled two sources with similar brightness in the Chandra image, while it is unresolved in the \xmm\ one. The image has been smoothed with a Gaussian kernel of radius 1 pixel.}
\label{fig:src901}
\end{figure}

In order to identify sources that were potentially blended in the \xmm\ data our ULX sample was initially selected from, we first examined cases with multiple source matches in CSC2 within our 10$''$ search radius. Of the 296 \xmm\ sources with multiple X-ray observations considered here, 34 returned multiple \chandra\ matches. In these instances, we retained the initial \xmm\ source if one of the matched sources was significantly brighter than the others (by an order of magnitude or more in flux), and was the closest match to the \xmm\ position, as this clearly identifies it as the real \chandra\ counterpart and implies that the \xmm\ data was dominated by a single source. However, in a number of cases the \chandra\ data revealed two (or more) sources of similar brightness within the \xmm\ PSF, implying the \xmm\ `source' was likely dominated by the combination of these sources in reality (for example, see Figure \ref{fig:src901}). We excluded these 18 sources from our sample.
There were 129 \xmm\ selected sources for which no \chandra\ data exist in CSC2, and so no further assessment of source confusion can be made beyond the initial selection in the parent ULX catalogue, which only included sources consistent with being point-like in the \xmm\ data (\citealt{Earnshaw19}).

\subsection{Initial Sample Selection}

It is important to note at this stage that the data compiled from \xmm, \swift\ and \chandra\ each cover slightly different energy bands. In order to combine the data into a single long-term lightcurve for each of the 278 remaining sources, we therefore converted the fluxes/limits to a common 0.3--10 keV energy band (where necessary) using \webpimms\ and the spectral models used to calculate the fluxes for each of the different missions. From these lightcurves, which utilized the FLIX upper limits for \xmm\ and the \swift\ data from the default first pass with the XRT pipeline, we initially selected 45 sources that, even after considering the statistical uncertainties on the fluxes, show long-term variability larger than an order of magnitude (selected to match the variability threshold used in \citealt{Earnshaw2018}). These sources were further inspected in greater detail to produce refined lightcurves (see below) and ensure the inferred level of variability was robust.

\subsection{\textit{XMM-Newton} Sanity Checks}

We began by inspecting in detail the \xmm\ data for our initial sample of 45 in order to confirm the veracity of the source detections, and refine the initial upper limits provided by FLIX in the case of non-detections. 

\subsubsection{Spurious Detections}

Although the 3XMM-DR7 catalogue used various methods to try and ensure robust source detections, it is still possible that some of the catalogue entries are actually spurious. We therefore inspected the \xmm\ images for each of our initial sample, and concluded that two entries were likely spurious, for a variety of reasons. SRC~90011 only has a single \xmm\ detection, which was just above the detection threshold for the 3XMM survey, and the source was in the wings of the PSF of the bright X-ray source corresponding to the central AGN in Mrk\,3. We further confirmed that it did not appear in the \chandra\ catalogue and was not detected in the \swift\ image. Source 354089 only had a single, low-significance \xmm\ detection, and was located right at the edge of the \xmm\ field-of-view (FoV). Again, there were no corresponding detections with either \chandra\ or \swift. Furthermore, SRC~354089 was seen in the direction of the nucleus of its host galaxy, NGC\,4151 (another extremely X-ray bright AGN), which was just outside of the \xmm\ FoV for the observation in which this source is detected. We therefore excluded these two sources from our final sample.

In addition, although the source itself is not spurious, we found that SRC~10388 and 3277 both had two entries for the same observation ID in 3XMM-DR7. These had slightly different source positions, and fluxes that differed by roughly an order of magnitude. We conservatively adopt the entry that had a higher detection significance, which gave a higher flux. SRC~3277 then did not meet our variability threshold when compared against the detections in other observations, and so was also excluded from our final sample.

\subsubsection{Source Misidentification}

Inspecting the \xmm\ images for sources 28943 and 349814, we realised these two sources are only separated by 4$''$. There were 4 \xmm\ observations covering NGC~4485, the host galaxy of these two source entries. Each source had two detections, and two upper limits. In both cases, these upper limits were the reason that the sources were initially selected as highly variable. However, the two upper limits for SRC~349814 corresponded to the two detections in SRC~28943 and vice versa. Furthermore, the \chandra\ images only showed a single source near the position of the two \xmm\ sources. We therefore concluded that 3XMM-DR7 incorrectly assigned two different IDs to the same source in this case, resulting in each only appearing to be detected in some of the available observations. After merging the measurements into a single lightcurve and reassessing the variability of the source, we found that it did not meet our variability threshold, and so we also excluded these data from our final sample. This is the only case in which such misidentification appears to have occurred within our sample.

\subsubsection{Other Imaging Issues}

\begin{figure}
\centering
\includegraphics[width=\linewidth]{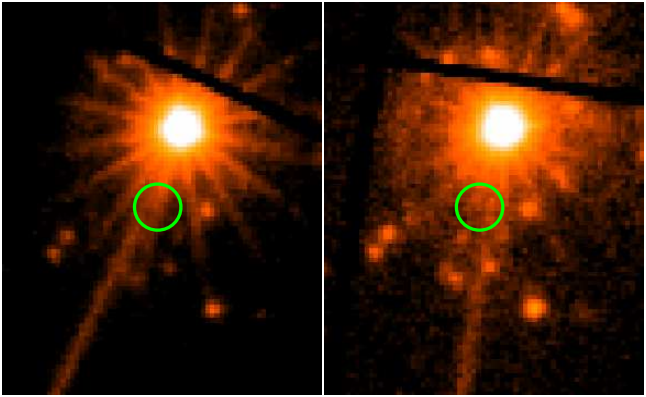}
\caption{\xmm\ PN images for SRC~17826 from obsID~0692840401 (left) and 0692840501 (right), both smoothed with a Gaussian kernel of radius 1 pixel. The source position is marked with a green solid circle of radius 20$''$. The source position in these two observations is affected by the CCD read-out column and the bright AGN.}
\label{fig:17826}
\end{figure}

During the course of inspecting the \xmm\ images, we also identified a small number of individual observations where the source in question was located in a sufficiently complex environment that good fluxes/upper limits could not be obtained. For example, for two observations of SRC~17826, in the galaxy NGC\,1365, the position of the ULX (which does not appear to have been detected) was located in both the wings of the PSF of the X-ray bright central nucleus and its readout streak (see Figure \ref{fig:17826}). This meant there was no available location sufficiently similar to the source position with which to perform a suitable local background estimate, and so we were unable to estimate a robust upper limit. These observations were excluded from the final lightcurve for this source. That these issues only resulted in the exclusion of a small number of observations, and in turn only resulted in one source being removed from the sample completely (SRC~8408).

\subsubsection{Upper Limits}
\label{sec:an_upl}

\begin{figure}
\centering
\includegraphics[width=\linewidth]{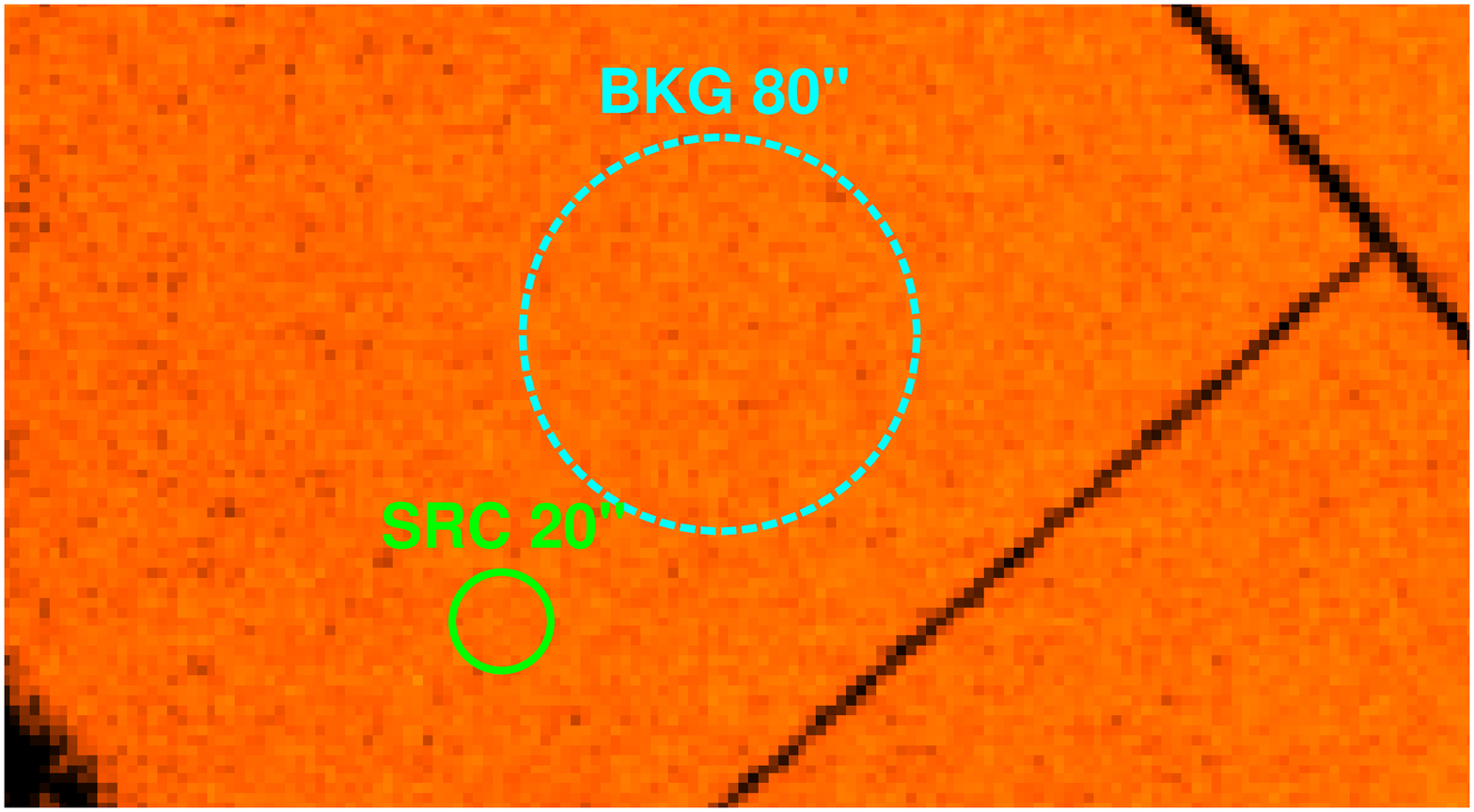}
\includegraphics[width=\linewidth]{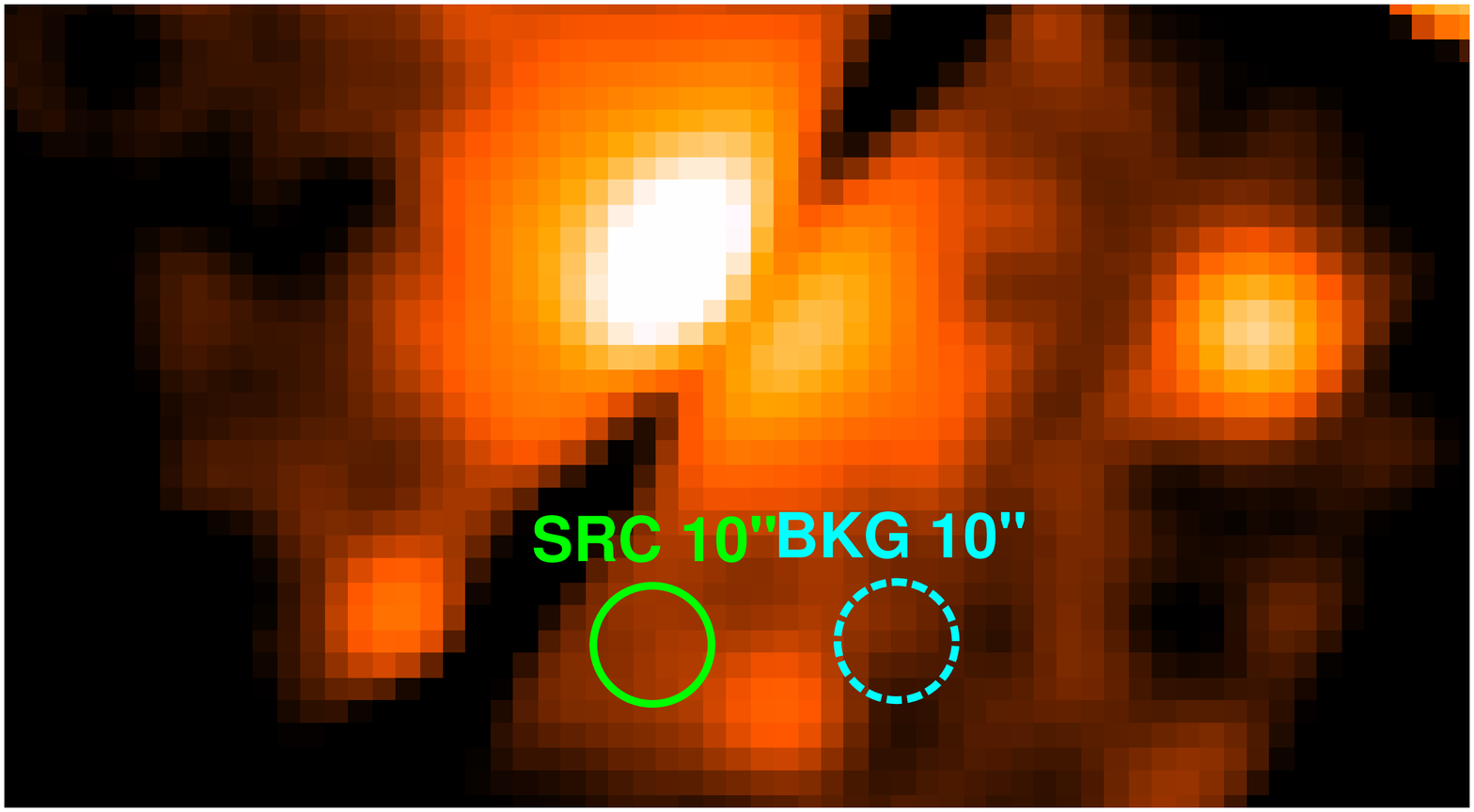}
\caption{Two examples of background region selection for \xmm\ observations. The upper panel shows the MOS2 image from obsID~0150280701 for SRC~4934, where the background is relatively clean and a large region is used. The lower panel shows the PN image for SRC~7245 from obsID~0303420201, where the background is chosen to have the same size as the source, since the source region is affected by the nearby bright source and the diffuse emission from the host galaxy. The source and background regions are marked in green solid and cyan dashed curves, respectively. The left and right figure are smoothed with a Gaussian kernel of radius 1 and 3 pixels, respectively.}
\label{fig:xmm_upl}
\end{figure}

As described in Section \ref{sec:upl}, the FLIX upper limits are likely to be underestimated. For each of the initially selected sample, we therefore manually re-calculated any \xmm\ upper limits in their lightcurves following the method outlined in Section \ref{sec:upl}. Figure \ref{fig:xmm_upl} shows two examples of the region selected for the upper limit assessment. For SRC~4934, where the source was not affected by other sources of emission, a large background radius of 80 arcseconds was selected. On the other hand, SRC~7245 (host galaxy M51) sits in the PSF of a nearby, bright source (the nucleus of M51), and also has another fainter source in close proximity. The background region is placed at approximately the same radial distance to both the nearby sources as the source region used for the upper limit calculation. In addition, this region also includes the diffuse emission from the host galaxy M51. As expected, the manually calculated upper limits (which we expect to be more robust) were larger than those obtained from FLIX. We then reassessed whether the sources would still meet our variability threshold. We identified three (SRC~27414, 35286, 64434) sources for which the FLIX upper limits were significantly underestimated, owing to the presence of significantly enhanced local backgrounds, and so the level of variability was significantly overestimated. These were therefore excluded from our final sample.

\subsection{\textit{Swift} Sanity Checks}
\label{sec:swift}

\begin{figure}
\centering
\includegraphics[width=\linewidth]{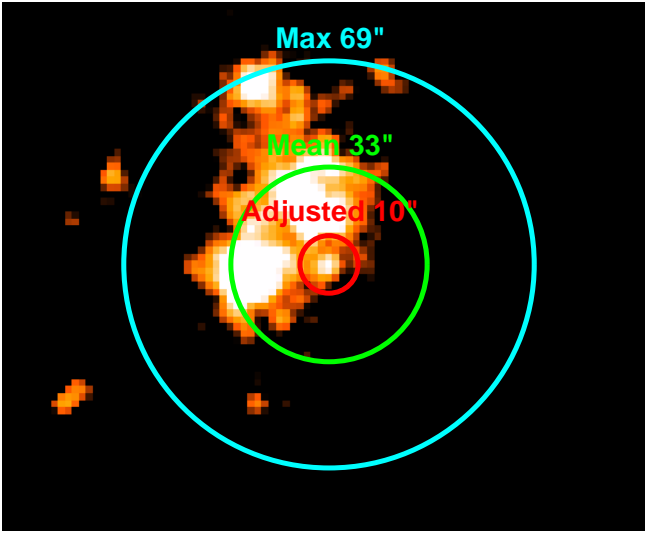}
\caption{\swift\ image of SRC~226383 at the source region, smoothed with a Gaussian kernel of radius 2 pixel. With centroiding turned off, the maximum region in green given by the Swift pipeline with radius of 69$''$ contains a few bright sources. The average 33$''$ source region in cyan is still contaminated by the nearby sources. After reducing the source region, the white circle shows the adjusted region centered at the \xmm\ position with radius of 10$''$.}
\label{fig:swift_226383}
\end{figure}

As noted above, the default XRT lightcurve pipeline utilizes a dynamic source region, which is designed to maximise the SNR for each individual \swift\ observation. The pipeline attempts to centroid the source region based on the first XRT image (in the case that the pipeline fails to centroid, the stacked image is used instead.), and adjusts the size of the extraction region on an observation-by-observation basis with radii limited to the range 11.8--70.8$''$ \citep{Evans2007}. However, a number of the ULXs considered here are in fairly crowded fields, and for these sources this approach can become problematic. In some cases, the presence of bright, nearby sources can confuse the centroiding process, resulting in offset source regions that may then also contain flux from both (or even multiple) sources. In other cases, even if the centroiding keeps the region centred on the source in question, for some observations it can still be advantageous for the pipeline to increase the size of the source region to incorporate additional flux from a nearby source as this increases the S/N of the integrated data within that region (for example, if a nearby transient source appears later in the \swift\ coverage). These issues primarily result in the \swift\ data overpredicting the true source flux; we show the example of SRC~226383 in Figure \ref{fig:swift_226383}, which suffers from contamination. 

For each of the remaining sources in our initial selection, we therefore compared the position of the source region and the maximum size used in the initial \swift\ analysis with the various X-ray images available. Where any of the issues highlighted above were observed, we re-ran the \swift\ pipeline either with centroiding turned off or with a manually defined maximum size for the source region (or both), where relevant. Where an upper limit on the size of the source region was set, this was determined on a case-by-case basis, depending on the proximity of the other nearby sources. After re-running the \swift\ pipeline with updated settings, we again re-analysed their levels of variability.

We draw special attention to SRC~17826, where the source position was significantly contaminated by the PSF of the nearby AGN (see Figure \ref{fig:17826} for the environment around this source). An examination of the \swift\ images did not show a source detection (Evans et al., in prep.). However, even when selecting a background region close to the source to compute the \swift\ fluxes and upper limit, we found that the pipeline returned excess fluxes at the source position. Thus, we treated the 3$\sigma$ upper bound of these fluxes as upper limits. This is the only case where a strong contamination was observed even after using carefully chosen source and background regions, resulting in unreliable flux measurements when using the \swift\ pipeline. Nevertheless, the source can still be classified as highly variable given the available \chandra\ and \xmm\ data. 

11 of our initial sample (SRC~3374, 28995, 40237, 44195, 55654, 122918, 226383, 348319, 358069, 359377, 366822) no longer met our variability threshold with the refined \swift\ data, and so were removed from our final sample.

\subsection{The Final Sample}
\label{sec:samples}

In summary, after addressing the identified issues above, we excluded 20 sources from our initial sample: 2 because of misidentification, 3 spurious \xmm\ sources, 1 due to a spurious \xmm\ flux measurement, 3 after upper limit refinements, and 11 based on the reprocessed Swift fluxes. We were therefore left with a final sample of 25 highly variable ULXs. These are listed in Table \ref{tab:var}, along with some of their basic properties. These sources revealed a variety of different long-term behaviour, which we discuss further below. For each of these sources we constructed a final long-term lightcurve, and also computed flux distributions based on these lightcurves (similar to \citealt{Tsygankov2016} and \citealt{Earnshaw2018}). For comparison with the rest of our highly variable sample, we showed the long-term lightcurve and the flux distribution for the known PULX NGC\,5907 ULX1 (which, as expected, is also selected by our analysis) in Figure \ref{fig:28355}, and we also show a few other individual sources from our sample in Figures \ref{fig:2242}, \ref{fig:100854} and \ref{fig:90213}; the remaining sources are shown in Appendix~\ref{ap:fig}.

One thing that is immediately apparent from these plots is the extremely variable coverage currently available for these sources. Although there were a number of cases that have been observed fairly frequently, there were also a number of cases with extremely sparse coverage. The poor coverage available for these sources prevented us from undertaking a systematic statistical analysis of these distributions to try and formally quantify any degree of bi-modality/deviation from standard behaviour for persistently accreting sources (see below). We therefore limited ourselves instead to a simple visual assessment to determine whether the sources selected show any evidence for off-states that could potentially be related to propeller transitions. These assessments are also given in Table \ref{tab:var}. In total, we found 17 new sources in our highly variable sample that either show good evidence for such off-states, or at least are consistent with doing so.

\begin{sidewaystable*}
%\vspace*{-15cm}
\caption[labelfont=bf]{Key properties for our sample of highly variable ULXs (i.e. those that exhibit long-term variability of more than a factor of 10). Max/Min shows the maximum factor of flux variability observed to date; if the minimum is an upper limit, this value actually indicates a lower limit to the maximum variability. The maximum fluxes are given for the 0.3--10 keV band, and the maximum luminosities are calculated by $F \times 4 \pi D^2$.}
\vspace*{-0.4cm}
\begin{threeparttable}
\resizebox{\linewidth}{!}{\begin{tabular}{ccccccccp{7cm}}
\hline \hline
3XMM DR4 & IAU Identifier & \multirow{2}*{Host Galaxy} & Common  & $D$ & Max/Min & Max Flux & Max Luminosity & \multirow{2}*{Comments} \\
SRCID & (3XMM ...) & & Name & (Mpc) & Flux &  (10$^{-13}$ erg cm$^{-2}$ s$^{-1}$) & (10$^{39}$ erg s$^{-1}$) & \\ \hline
2242\tnote{$\ast$} & J203500.1+600908 & NGC 6946 & ULX1\tnote{a} & 7 & $>$18 & 8.22$\pm$1.36 & 4.50$\pm$0.74 & Roughly log-normal \\
4934 & J032004.9-664211 & NGC 1313A &  & 74 & $>$20 & 3.62$\pm$1.42 & 238.52$\pm$93.52 & Some evidence for bi-modality, but could be log-normal given the similar luminosity of the two peaks \\
5256 & J133000.9+471343 & NGC 5195 & ULX3\tnote{a}, ULX7\tnote{d} & 9 & 163 & 9.69$\pm$1.30 & 9.60$\pm$1.29 & Known PULX, clear evidence for off-states \\
7245\tnote{$\ast$} & J132953.3+471042 & NGC 5194 & ULX4\tnote{a,d} & 9 & 191 & 3.76$\pm$0.17 & 3.72$\pm$0.16 & Good evidence for bi-modality/off-states \\
10388 & J034615.7+681112 & IC 342 & X-2\tnote{b} & 3 & 11 & 111.58$\pm$0.95 & 14.85$\pm$0.13 & Skewed towards higher luminosity but no evidence for off-states \\
16647 & J181943.4+743336 & NGC 6643 &  & 20 & $>$13 & 1.02$\pm$0.29 & 4.78$\pm$1.35 & Roughly log-normal, but shows a potential off-state \\
17826 & J033337.9-360935 & NGC 1365 & X18\tnote{c} & 18 & $>$22 & 0.65$\pm$0.07 & 2.65$\pm$0.27 & Good evidence for bi-modality/off-states, but limited sensitive coverage \\
19949 & J122204.3+281110 & IC 3212 &  & 101 & 18 & 8.94$\pm$5.08 & 1100.47$\pm$625.07 & Consistent with bi-modality/off-states, but limited sensitive coverage \\
28355\tnote{$\ast$} & J151558.6+561810 & NGC 5907 & ULX1 & 17 & $>$177 & 20.66$\pm$5.06 & 71.05$\pm$17.41 & Known PULX, clear evidence for off-states \\
28744 & J122903.4+135816 & NGC 4459 &  & 16 & $>$21 & 0.55$\pm$0.05 & 1.67$\pm$0.16 & Consistent with bi-modality/off-states, but very sparse coverage \\
29687\tnote{$\ast$} & J230457.6+122028 & NGC 7479 &  & 32 & 11 & 4.89$\pm$0.22 & 59.01$\pm$2.69 & Roughly log-normal \\
29790 & J013651.1+154546 & NGC 628 & ULX1\tnote{d}  & 10 & 37 & 4.62$\pm$0.94 & 5.69$\pm$1.16 & Good evidence for bi-modality/off-states \\
87497 & J072719.6+854632 & NGC 2276 &  & 32 & 18 & 1.00$\pm$0.37 & 12.37$\pm$4.59 & Consistent with bi-modality/off-states, but very sparse coverage \\
87501 & J072722.2+854513 & NGC 2276 &  & 32 & $>$35 & 2.15$\pm$0.66 & 26.66$\pm$8.20 & Consistent with bi-modality/off-states, but very sparse coverage \\
90213 & J073650.0+653603 & NGC 2403 &  & 3 & $>$456 & 7.95$\pm$0.35 & 0.99$\pm$0.04 & Good evidence for bi-modality/off-states \\
100854 & J013636.4+155036 & NGC 628 & ULX2\tnote{d} & 10 & $>$138 & 2.12$\pm$0.13 & 2.62$\pm$0.15 & Single-detection \\
102935 & J022233.4+422026 & NGC 891 & ULX1\tnote{e} & 9 & $>$2107 & 39.97$\pm$9.82 & 41.46$\pm$10.19 & Good evidence for bi-modality/off-states \\
266604 & J213631.9-543357 & NGC 7090 &  & 9 & $>$67 & 6.04$\pm$0.28 & 5.50$\pm$0.26 & Consistent with bi-modality/off-states, but fairly sparse coverage \\
279969 & J102957.2-351420 & NGC 3269 &  & 50 & 15 & 1.36$\pm$0.47 & 40.64$\pm$14.19 & Consistent with bi-modality/off-states, but very sparse coverage \\
326570 & J140338.4-335753 & NGC 5419 &  & 55 & 10 & 2.57$\pm$1.02 & 93.11$\pm$37.08 & Consistent with bi-modality/off-states, but sparse coverage \\
346790 & J143235.6-441003 & NGC 5643 &  & 16 & $>$114 & 2.81$\pm$0.93 & 8.61$\pm$2.85 & Consistent with bi-modality/off-states, but limited sensitive coverage \\
352909 & J123558.4+275741 & NGC 4559 & ULX1\tnote{a} & 10 & $>$12 & 18.22$\pm$1.68 & 20.52$\pm$1.90 & Roughly log-normal but shows potential off-states \\
358052 & J121847.6+472054 & NGC 4258 & XMM1\tnote{f} & 8 & $>$117 & 2.94$\pm$0.14 & 1.99$\pm$0.10 & Single-detection \\
358229 & J121920.8+055104 & NGC 4261 &  & 29 & $>$14 & 0.28$\pm$0.04 & 2.95$\pm$0.40 & Single-detection \\
366059 & J124820.6+082919 & NGC 4698 &  & 13 & 16 & 0.69$\pm$0.25 & 1.50$\pm$0.54 & Consistent with bi-modality/off-states, but very sparse coverage \\ \hline
\end{tabular}}
\begin{tablenotes}[para]
\item[$\ast$] Sources that have multiple \chandra\ source matches within 10$''$ of the \xmm\ source position. \\ \item[a] \citealt{LiuB2005} \item[b] \citealt{Bauer2003} \item[c] \citealt{Strateva2009} \item[d] \citealt{LiuM2005} \item[e] \citealt{HodgesKluck12} \item[f] \citealt{Winter2006} 
\end{tablenotes}
\end{threeparttable}
\label{tab:var}
\end{sidewaystable*}

\section{Discussion}
\label{sec:discu}

Motivated by the recent discovery of ULXs pulsars, we have undertaken a program to identify additional PULX candidates by searching for ULXs that exhibit strong long-term variability, and in particular low-flux `off'-states, which have been observed in the known PULXs and may be linked to the propeller effect (\citealt{Tsygankov2016}). Our work built on the initial search presented in \cite{Earnshaw2018} by undertaking a more comprehensive analysis of the available data in the archive. To undertake our search, we constructed the long-term X-ray lightcurves of 278 ULXs using all available observations from \xmm, \swift\ and \chandra. Because of the generally limited number of observations of a given source with each individual observatory, combining the data from different telescopes increases the chance of finding ULXs that exhibit such variability. We found 25 sources show long-term flux variability in excess of an order magnitude. Among this sample, we identified 17 new sources that could potentially exhibit a bi-modal flux distributions or off-states, similar to the known PULXs.

\subsection{Examples of Highly Variable ULXs}

\begin{figure*}
    \centering
    \includegraphics{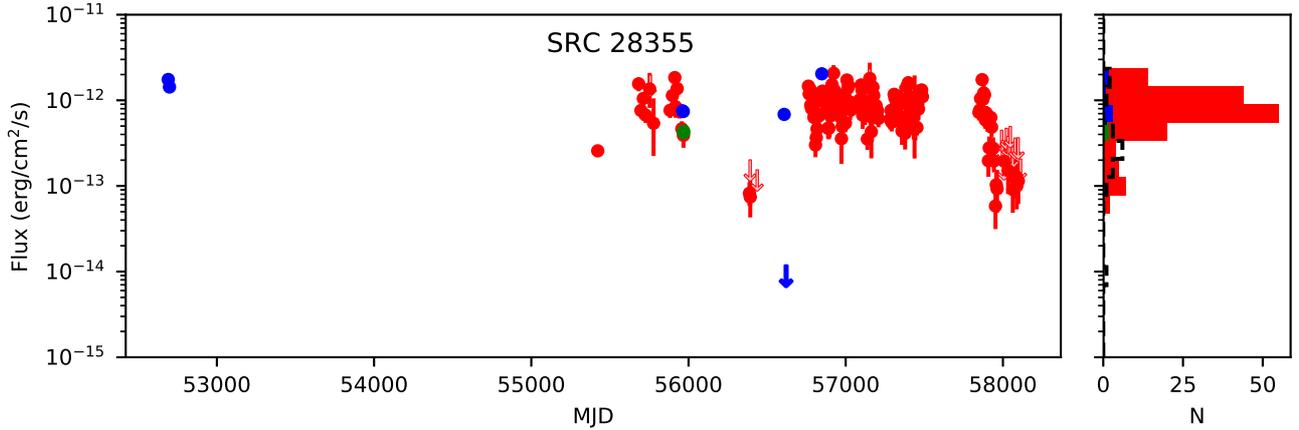}
    \caption{Long-term lightcurve and flux distribution for SRC~28355 (NGC\,5907 ULX1, a known PULX). The X-axes of the lightcurve and the histogram panels are the observation time in MJD and the number of observations, respectively. The common Y-axis is the observed flux. The \xmm, \swift\ and \chandra\ fluxes/upper limits are marked in blue, red and green points/downward arrows, respectively. For the histogram, all fluxes are stacked together to give the overall distribution, and the upper limits are added together and plotted in black dash line.}
    \label{fig:28355}
\end{figure*}

In the following sections, we present some examples of the lightcurves and the flux histograms of ULXs from our highly variable sample to demonstrate the different types of variability observed. As a further sanity check, we note that the known PULX NGC 5907 ULX1 (SRC~28355) and M51 ULX7 (SRC~5256) are selected by our analysis (as expected), showing a variability of more than a factor of 100 and good evidence for a bi-modal flux distribution. The lightcurve and the histogram of NGC 5907 ULX1 are presented in Figure \ref{fig:28355} as an example. Our approach confirms that PULXs can exhibit large flux variability and the analysis we have done is reasonable. 

\subsubsection{`Normal' Sources}

\begin{figure*}
    \centering
    \includegraphics{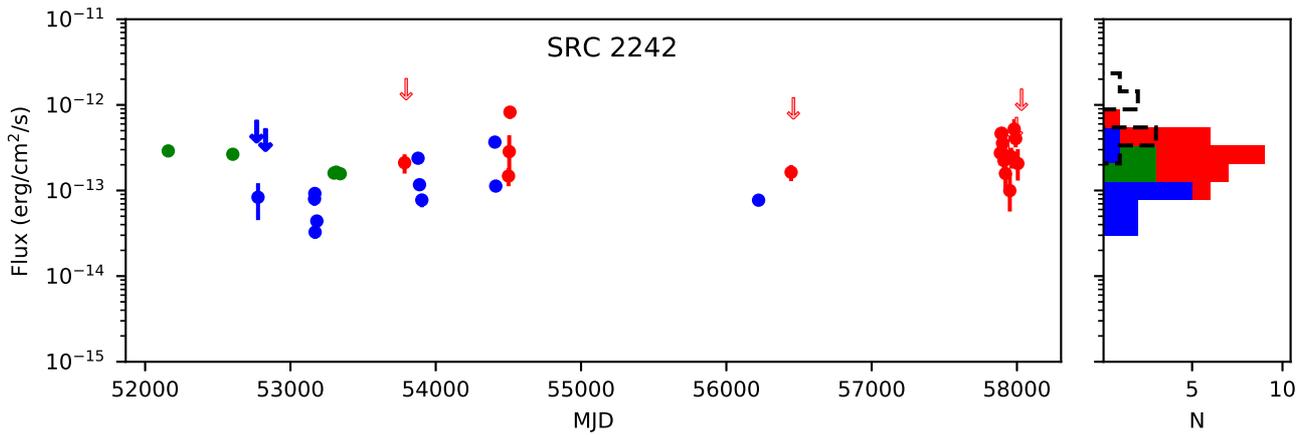}
    \caption{Lightcurve and histogram of SRC~2242 (NGC\,6946 ULX1). See caption of Figure \ref{fig:28355}.}
    \label{fig:2242}
\end{figure*}

There are a number of sources that exhibit observed variability amplitudes large enough to match our selection criterion, but broadly appear to show a continuous flux distribution, with no evidence for distinct off-states. In many cases, these distributions appear to be consistent with being approximately log-normal, as expected for accretion processes (\eg\ \citealt{Uttley05}). Figure \ref{fig:2242} shows the lightcurve and the histogram of SRC~2242 as an example of one such source. While this is the behaviour that would broadly be expected for BH ULX candidates, as these sources cannot experience propeller transitions, it is also possible that these sources are powered by NSs that just do not enter the propeller regime. This possibility will be discussed further in Section \ref{sec:imp}.

\subsubsection{Single Detections}

\begin{figure*}
    \centering
    \includegraphics{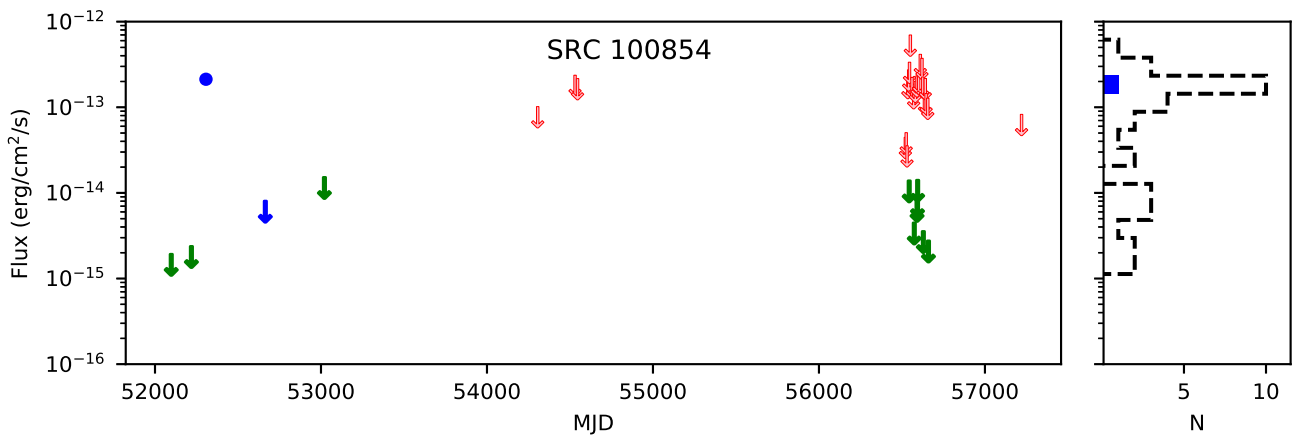}
    \caption{Lightcurve and histogram of SRC~100854. See caption of Figure \ref{fig:28355}.}
    \label{fig:100854}
\end{figure*}

There are also a couple of cases in which the source is mostly undetected, apart from a single observation that shows a flux higher than the ULX threshold. One such example is SRC~100854, shown in Figure \ref{fig:100854}. It can be seen that the lowest upper limits were well-separated with the detected flux. This behaviour could be explained by the presence of a transient ULX with a low duty-cycle. Indeed, the fourth PULX NGC 300 ULX1 is a transient system \citep{Carpano2018}, having been first detected in 2010 (when it was misidentified as a supernova and given the classification SN 2010da; \citealt{Monard2010}). However, such single detections could also be the result of explosive transient events (\eg\ genuine supernovae), which would naturally be short-lived, one-off events, and could well appear as a single detection given the limited coverage many of these candidates currently have. This `outburst' scenario is also seen in low mass X-ray binaries \citep[see \eg][]{Burke2013,Middleton2013}, where the source mostly stays in the sub-Eddington regime, but may reach ULX luminosity during outbursts. If the recurrence timescale is long and/or the sampling is sparse, such outbursts could result in just a single detection in our lightcurves. The detection of further outbursts from these sources is required to fully determine whether these sources are genuine, accretion-powered X-ray binaries.

\subsubsection{Potential PULX Candidates}

\begin{figure*}
    \centering
    \includegraphics{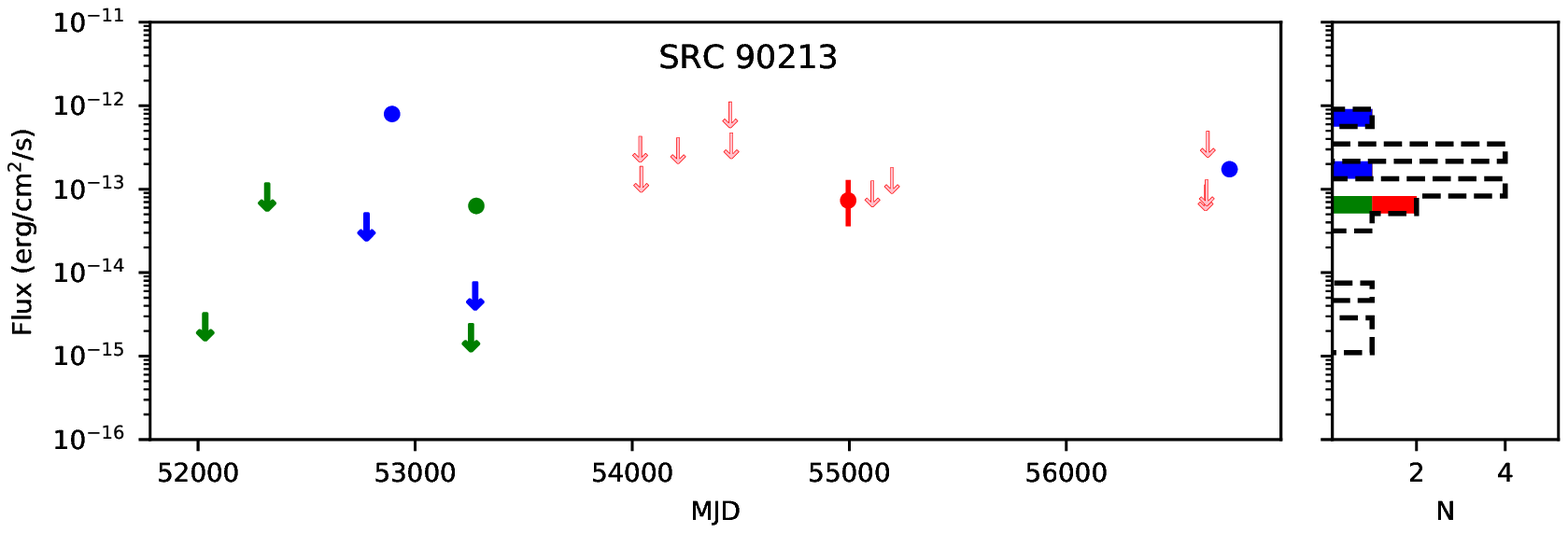}
    \caption{Lightcurve and histogram of SRC~90213. See caption of Figure \ref{fig:28355}.}
    \label{fig:90213}
\end{figure*}

The best PULX candidates among our sample are those that show good evidence for a bi-modal flux distribution, as would be expected for sources undergoing propeller phase transitions. A number of sources in our sample either do show good evidence for a bi-modal distribution, or at least are consistent with showing a bi-modal distribution within the limited coverage currently available. We show one of the best examples, SRC~90213 in Figure \ref{fig:90213}. Its long-term behaviour can be compared with NGC\,5907 ULX1 (shown in Figure \ref{fig:28355}) and NGC\,7793 P13, showing a high-state with ULX luminosities and a low-state orders of magnitude lower in flux. Among our highly variable sample, these sources are likely the highest priority in terms of continued monitoring to confirm their bi-modal nature, particularly given the sparse coverage currently available for a number of them.

Among our new bi-modal candidates (i.e. sources that are not already known to be pulsars), we note in particular that sources 90213, 102935 (NGC\,891 ULX; \citealt{HodgesKluck12}) and 266604 (NGC\,7090 ULX) have comparable peak fluxes to NGC\,5907 ULX1, currently the faintest (in terms of observed flux) of the known PULXs. It may therefore be possible to undertake meaningful pulsation searches for these sources with our current X-ray facilities. Furthermore, based on the available coverage, source 102935 appear to spend the majority of the time in their high-flux states, in which pulsation searches can most efficiently be performed. We also note that, as expected, our analysis additionally selected source 7245 (M51 ULX4), which is the bi-modal source highlighted by \cite{Earnshaw2018}. Although \cite{Earnshaw2018} did not detect any coherent pulsations from the data currently available for this source, we note that its peak flux is rather low in comparison to all of the known PULXs (roughly a factor of 5 fainter).

\subsection{Implications}
\label{sec:imp}

The 17 sources identified as showing evidence for a bi-modal flux distribution are our strongest PULX candidates, and as such are good targets for deeper follow-up observations to search for pulsations, either with current or future X-ray facilities. There are also a number of cases that are consistent with being bi-modal, but currently have poor coverage, so targeted monitoring of these sources to more robustly determine their flux distributions would also be particularly useful. 

In addition to helping to determine the contribution of neutron stars to the broader ULX population, which is currently a subject of significant debate (\eg\ \citealt{Pintore2017, Middleton17, Koliopanos17, Wiktorowicz17, Walton2018}), the identification of additional PULXs is an important step in understanding just how these remarkable sources are able to reach such extreme apparent luminosities. In particular, if the off-states that we have used to select our PULX candidates are associated with propeller transitions, this offers a potential means to estimate the magnetic fields of these systems. This is currently another area of significant debate (\eg\ \citealt{Eksi2015, DallOsso2015, King2016}), but is a key quantity in terms of determining accretion physics for these systems. For sources to undergo such a transition the magnetospheric radius ($R_{\rm{m}}$) must be similar to the co-rotation radius ($R_{\rm{co}}$). In the standard model for magnetically dominated accretion (\citealt{Ghosh1977}), $R_{\rm{m}}$ is determined by both the magnetic field ($B$-field) and the mass accretion rate ($\dot{M}$, which should itself be related to the observed flux): $R_{\rm{m}} \propto B^{4/7} \dot{M}^{-2/7}$, while $R_{\rm{co}}$ is determined by the spin period of the neutron star ($P$): $R_{\rm{co}} \propto P^{2/3}$. While we do not have $R_{\rm{co}}$ at the current time for these sources, if pulsations are identified in the future, knowing $R_{\rm{co}}$ helps the estimation of $R_{\rm{m}}$ and thus the strength of the $B$-field (although it should be noted that the $B$-field measured this way may only probe the dipolar component, and would not necessarily shed light on any higher-order components to the overall magnetic field that act closer to the neutron star, \eg\ \citealt{Israel2017}).

For the other sources highlighted here, which have observed variability amplitudes larger than an order of magnitude but do not show good evidence for off-states (i.e. they show a more `normal' flux distribution), they could still be NSs that do not enter the propeller regime. If this is the case, then this would likely imply that $R_{\rm{m}} \ll R_{\rm{co}}$, such that even an order of magnitude variation in flux is not sufficient to trigger a propeller transition. In turn, this would then imply that these sources have weaker $B$-fields or larger spin-periods (or both), when compared to sources that have similar peak luminosities but do undergo propeller transitions. For sources with $R_{\rm{m}} \ll R_{\rm{co}}$ we would expect the accretion disk to make a stronger relative contribution to the total observed flux, which in turn makes the pulsations more challenging to detect (as the disk components should not pulse). This is qualitatively consistent with the broadband spectral analysis comparing the known PULXs with ULXs from which pulsations have not currently been seen (\citealt{Walton2018}). Finally, although we have focused on highly variable ULXs in this work, we also note that sources with more modest variability could again be NSs with $R_{\rm{m}}$ always smaller than $R_{\rm{co}}$. However, since their observed variability amplitudes are lower, the expected changes in $R_{\rm{m}}$ are subsequently smaller, and so the degree to which $R_{\rm{m}}$ would have to be smaller than $R_{\rm{co}}$ is correspondingly not as strongly determined.

\subsection{X-ray Colours}

\begin{figure}
    \centering
    \includegraphics[width=\linewidth]{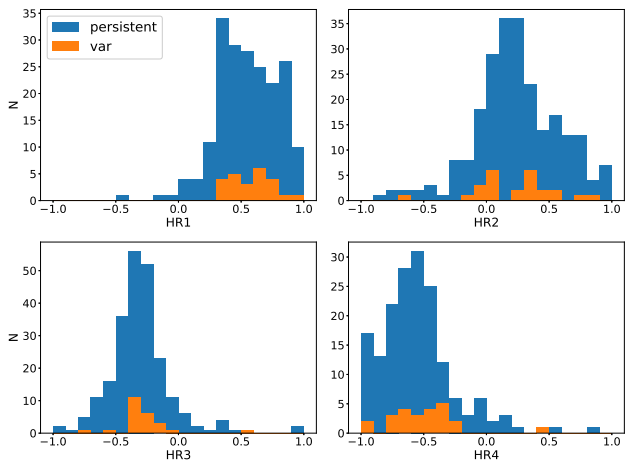}
    \caption{Histograms of the four HRs ratios for the highly variable sample (in orange, marked as `var') and the rest of ULXs (in blue, marked as `persistent'). The histogram has a bin size of 0.1.}
    \label{fig:hrs}
\end{figure}

We also investigated the X-ray colours/hardness ratios of our highly variable sample, and compared them to those of the broader ULX population to see if there are any notable spectral differences that could potentially be used to help identify other highly variable sources that do not currently have sufficient temporal coverage. We limited ourselves to a simple colour-based analysis given the highly variable data quality available, and focused on the \xmm\ hardness ratios since our parent sample is derived from the 3XMM catalogue. This provides source information for five sub-bands across the full 0.2--12.0\,keV \xmm\ energy range (0.2--0.5, 0.5--1.0, 1.0--2.0, 2.0--4.5 and 4.5--12.0\,keV), as well as four hardness ratios between adjacent sub-bands. These hardness ratios are defined as $HR = (H-S)/(H+S)$, where $H$ and $S$ are the count rates in the harder and the softer band, respectively (such that they are bounded by the range $-1 \leq HR \leq 1$). For simplicity, we focused on the average hardness ratios for each individual source (also provided by 3XMM for sources with multiple XMM detections), and computed histograms for each of the four hardness ratios for our highly variable sample and the rest of the ULX candidates included in the \cite{Earnshaw19} catalogue; we also followed the approach of \cite{Earnshaw19} and only included sources with robustly constrained hardness ratios (those with uncertainties less than 0.2). These histograms are shown in Figure \ref{fig:hrs}. We found that there is little to distinguish the highly variable sources from the rest of the ULX population with this simple analysis; in all cases the distributions of the two populations are clearly similar. This further stresses the need for continued monitoring programs to unearth more of these highly variable ULXs, as there does not appear to be a simple way to distinguish them based on their spectral properties.

\subsection{Limitations}
\label{sec:issues}

Although we have taken a number of steps to ensure the lightcurves produced are robust measures of the variations exhibited by the ULXs considered, there are a number of issues that could still potentially influence the variability amplitudes inferred. As discussed above, the majority of the \chandra\ fluxes used (from CSC2) were computed assuming a slightly different spectral model than assumed for the \xmm\ data in 3XMM and for the \swift\ data, potentially introducing a systematic offset between the fluxes inferred from these observatories. In addition, there are known cross-calibration issues between \chandra, \xmm\ and \swift\ which we have not actively accounted for in our work. However, both of these effects are at the $\sim$10--20\% level (see \citealt{Madsen15, Madsen17} for a recent assessments of the cross-calibration between these observatories). Since we have been searching for variations in excess of an order of magnitude, we consider it unlikely that these issues would have a significant effect on our assessment of the source variability. Additionally, our initial selection of highly variable sources was based on upper limits calculated by the FLIX server, which are likely not as robust as the manual calculations we subsequently performed for these sources. However, since the FLIX limits are likely underestimated, such that they would overestimate the variability amplitude, it is unlikely that using these limits for our initial selection would have caused us to incorrectly exclude any sources from our final sample.

It is also worth noting that the selection criterion used here (i.e. at least a factor of 10 in long-timescale flux variability) is purely empirical, and based on the observed behavior of PULXs. The expected difference in flux across the propeller transition can be expressed approximately as $\Delta L_{\rm{X}} \sim 170 P^{2/3} M_{1.4}^{1/3} R_{6}^{-1}$ (where $P$ is the spin period in seconds, $M_{1.4}$ is the neutron star mass in units of 1.4\,\msun, and $R_{6}$ is the neutron star radius in units of $10^{6}$\,cm, \citealt{Tsygankov2016}). If there are PULXs with spin periods significantly shorter than those seen to date (e.g. millisecond pulsars), the expected level of variability is smaller than our selection criterion. These sources will likely not be included in our sample, and in general will be difficult to identify among the broader ULX population from their long-term variability.

Another potential explanation for strong long-term flux variability is via the super-orbital periodicity seen in some PULXs, which can in some cases reach amplitudes similar to those selected here (\eg\ \citealt{Brightman2019}). While the origin of these cycles is not entirely clear, super-orbital periods are typically interpreted as being related to some kind of precession, rather than variations in accretion rate (\eg\ \citealt{Kotze12}). However, the majority of systems with robustly confirmed long-timescale X-ray periods are also known PULXs (\citealt{Walton2016, Fuerst18}; Brightman et al.\ 2019b, in prep.), so selecting sources with high-amplitude variability is still likely a reasonable way of identifying good PULX candidates, even if we are really seeing super-orbital variability in some cases.

However, the primary limitation to our work is the sparse coverage available for the majority of the ULXs considered. As discussed above, this prevents us from undertaking a more rigorous statistical analysis of the available flux distributions to test for bi-modality, and so we limited ourselves to a visual assessment, which is naturally more subjective. However, even more fundamentally, this lowers the probability of having observed off-states in many of these sources in the first place, even if intrinsically they do exhibit this behaviour. These issues can only be addressed with higher continued (and higher cadence) monitoring of a larger sample of ULXs. As discussed by \cite{Earnshaw2018}, the \erosita\ all-sky survey (\citealt{EROSITA_tmp}) will naturally provide additional coverage of all of these sources, and has good potential for discovering even more highly variable ULXs.

\section{Conclusions and Future Work}
\label{sec:con}

With the current sample of known PULXs still severely limited, identification of further members of this population is a critical step in our efforts to understand these enigmatic sources. One possible way to identify good PULX candidates among the broader ULX population, based on the behaviour seen from the known PULXs, is to search for sources exhibiting low-flux states in addition to their extreme ULX luminosities. These may be related to propeller transitions, which would require a neutron star accretor. Building on an initial search for such sources based on \xmm\ (\citealt{Earnshaw2018}), we compiled the available data from each of the \xmm, \swift\ and \chandra\ observatories for the sample of ULXs compiled by \cite{Earnshaw19}, and construct long-term lightcurves for each of these sources. Because we were looking for faint states, in which the source may not be detected, where this appears to be the case we paid particular attention to computing robust upper limits to the source flux so that we can accurately determine the amplitudes of the variability exhibited. Of the 278 ULX candidates with multiple observations, we identified 25 sources that showed at least an order of magnitude in variability. Among these 25, there are 17 new sources that appear to show off-states/bi-modal flux distributions similar to the known PULXs. 

These sources are good candidates for both continued monitoring (as a number have sparse coverage) and deeper follow-up observations that could help to identify pulsations. Pulsation searches for these sources are important for both expanding the sample of know PULXs, and confirming our approach as an efficient method of identifying PULXs among the broader ULX population. However, while some of the sources identified are bright enough for sensitive pulsation searches with our current X-ray observatories, many are faint and such work may require observations with the next generation of X-ray observatories (\eg\ \athena; \citealt{ATHENA}, and \erosita; \citealt{EROSITA_tmp}). Further expansion of the known ULX population in the local universe, combining continued monitoring, updated galaxy and X-ray source catalogues, would also potentially help to identify further examples of this behaviour.

\section*{Acknowledgements}
XS thanks the further support of the IoA Summer School Programme. DJW acknowledges financial support from STFC in the form of an Ernest Rutherford Fellowship. PAE acknowledges UKSA support. This research has made use of data obtained from the Chandra Source Catalog, provided by the Chandra X-ray Center (CXC) as part of the Chandra Data Archive. 

\appendix
\section{Long-Term Lightcurves}
\label{ap:fig}

\begin{figure*}
\centering
    \includegraphics{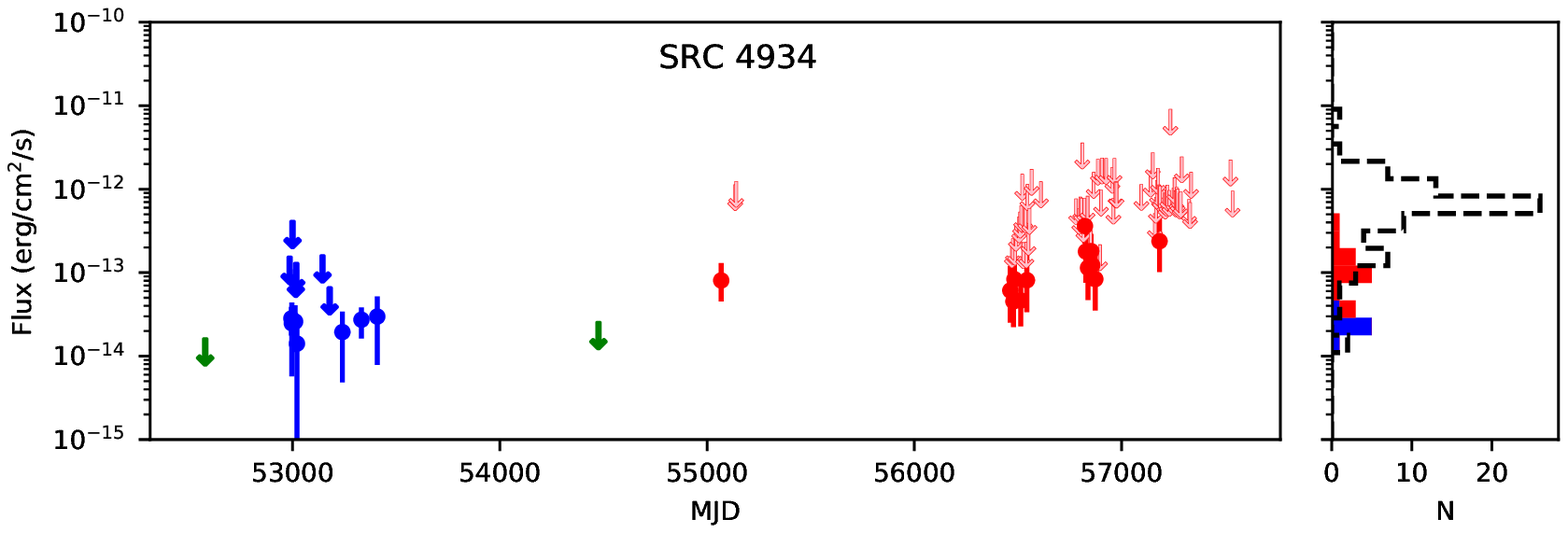}
    \includegraphics{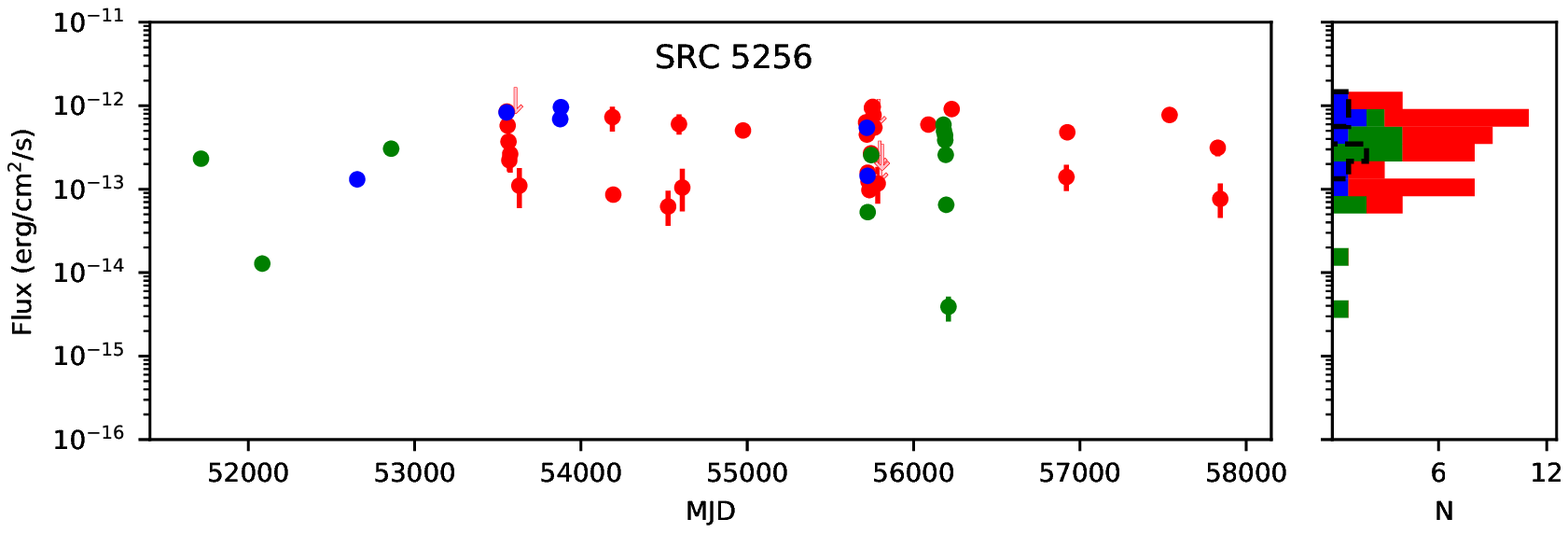}
    \includegraphics{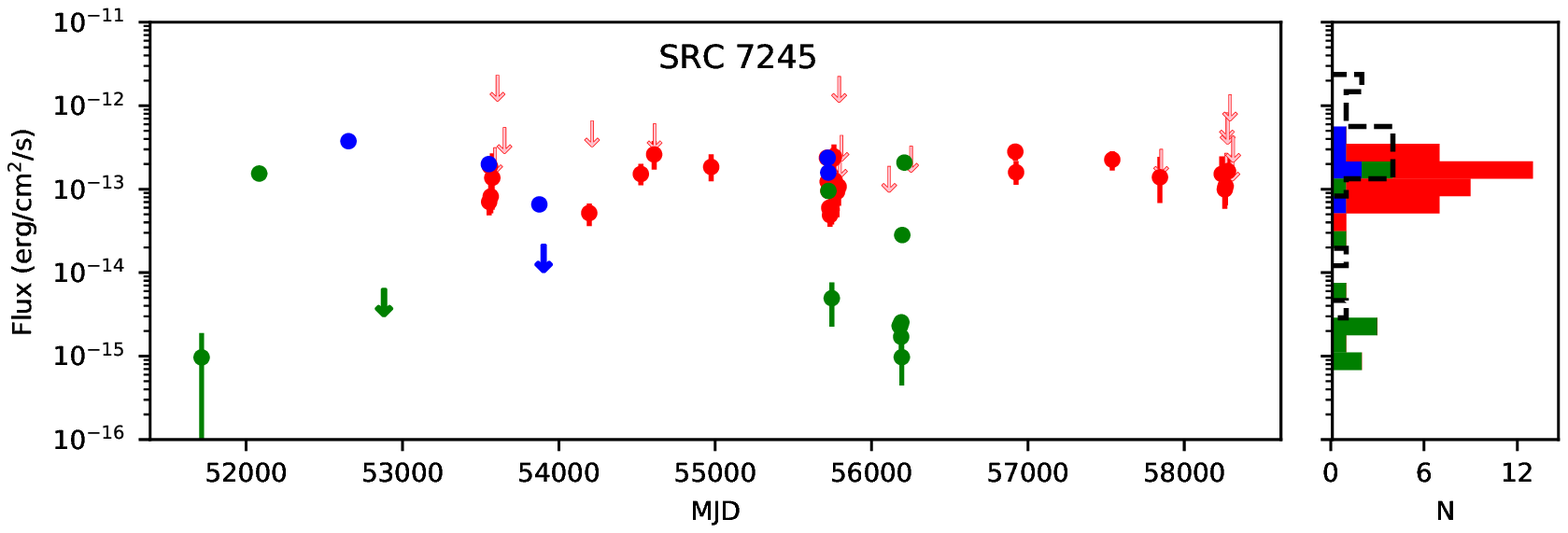}
    \includegraphics{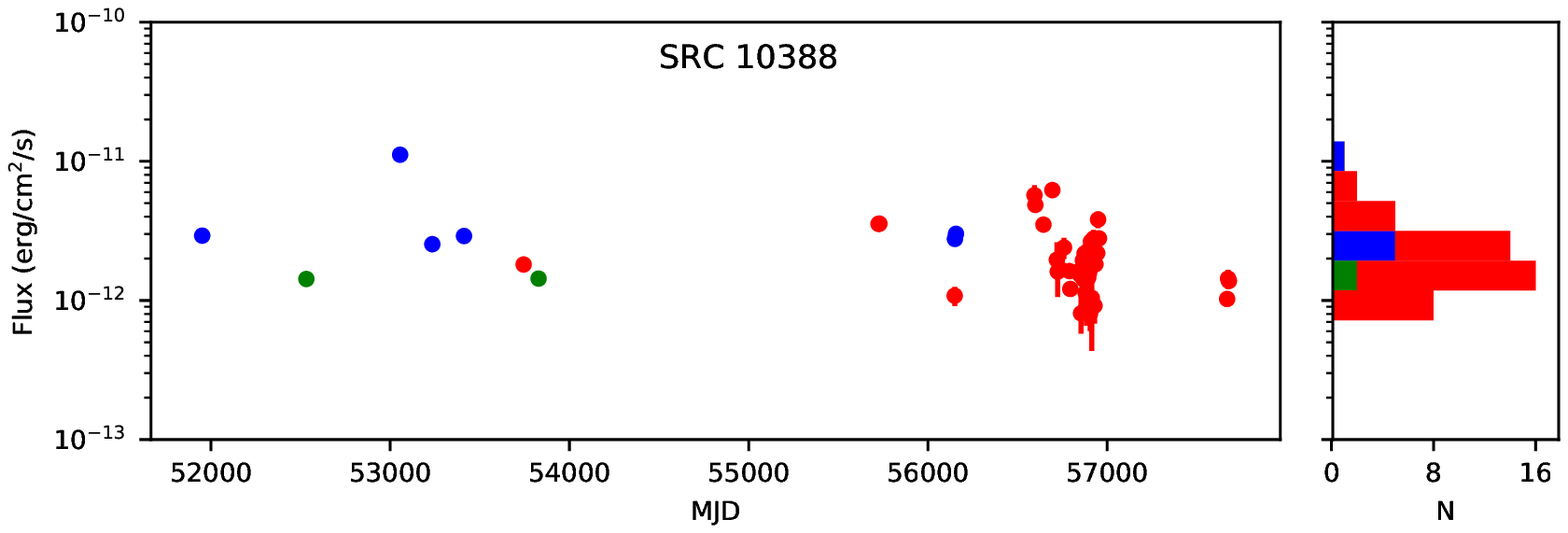}
    \caption{Lightcurves and flux histograms for the rest of our highly variable ULX sample (similar to Figures \ref{fig:28355}, \ref{fig:2242}, \ref{fig:100854} and \ref{fig:90213}).}
    \label{fig:sample_lcs}
\end{figure*}

Here we show the remaining long-term lightcurves compiled for our highly variable sample of ULX candidates (Figure \ref{fig:sample_lcs}).

\begin{figure*}
\centering
    \includegraphics{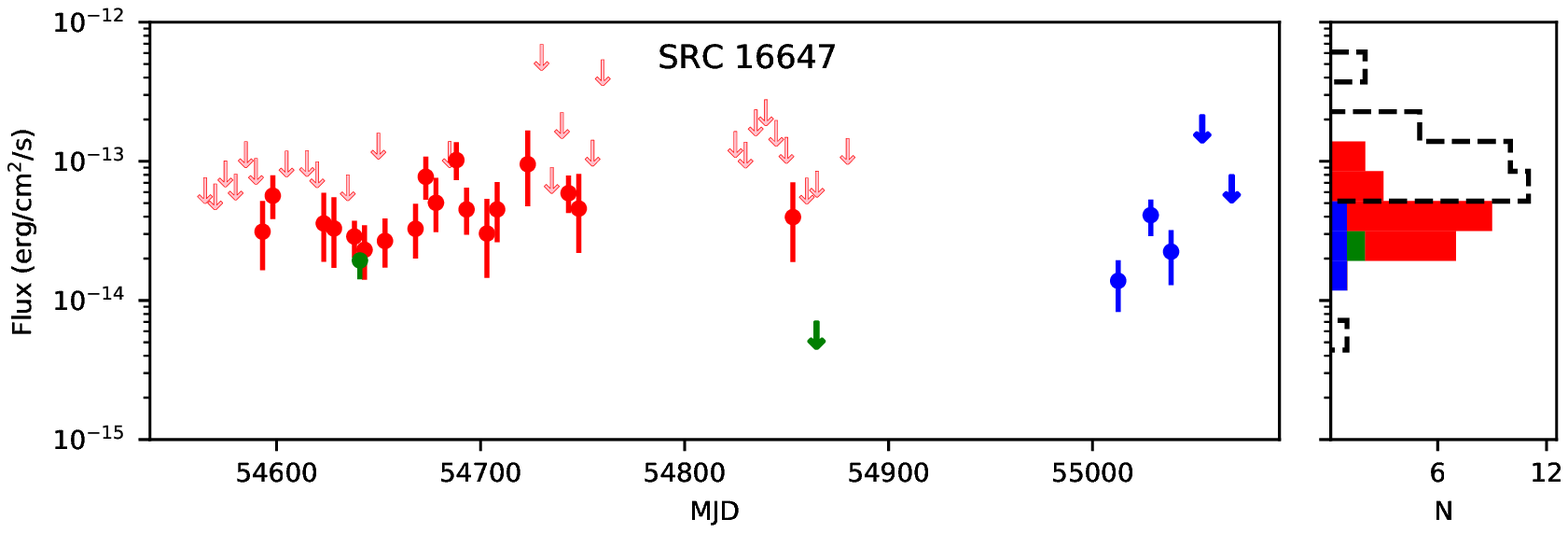}
    \includegraphics{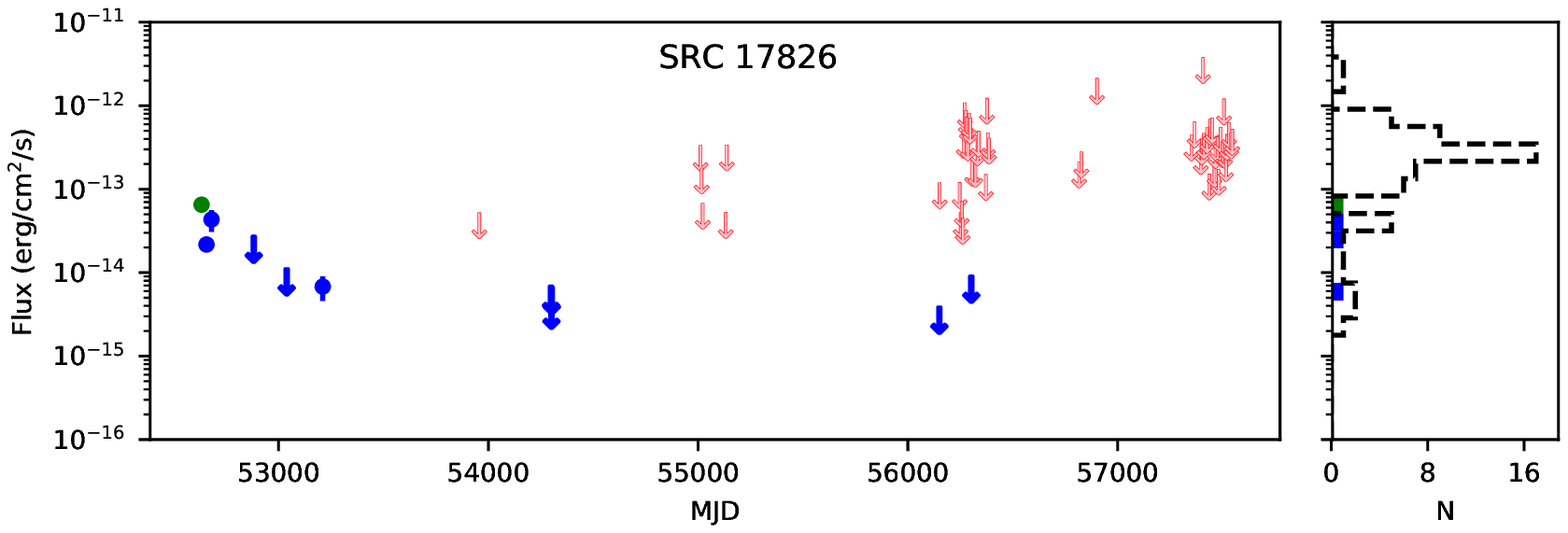}
    \includegraphics{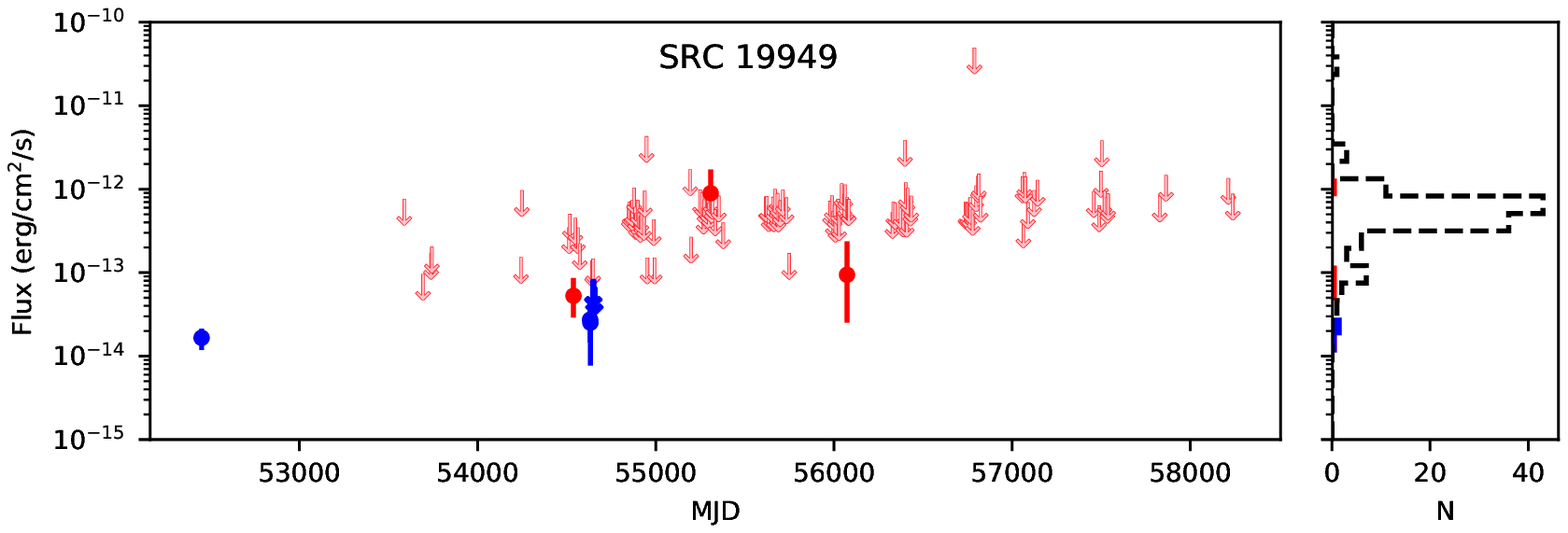}
    \includegraphics{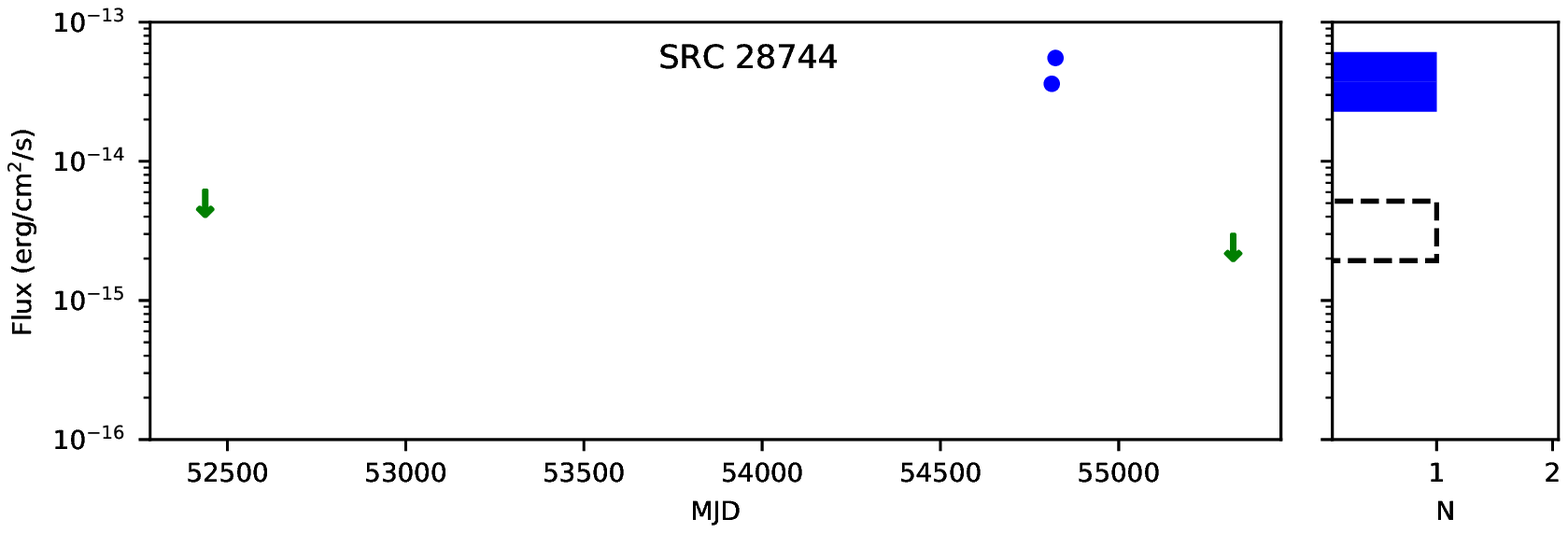}
\end{figure*}

\begin{figure*}
\centering
    \includegraphics{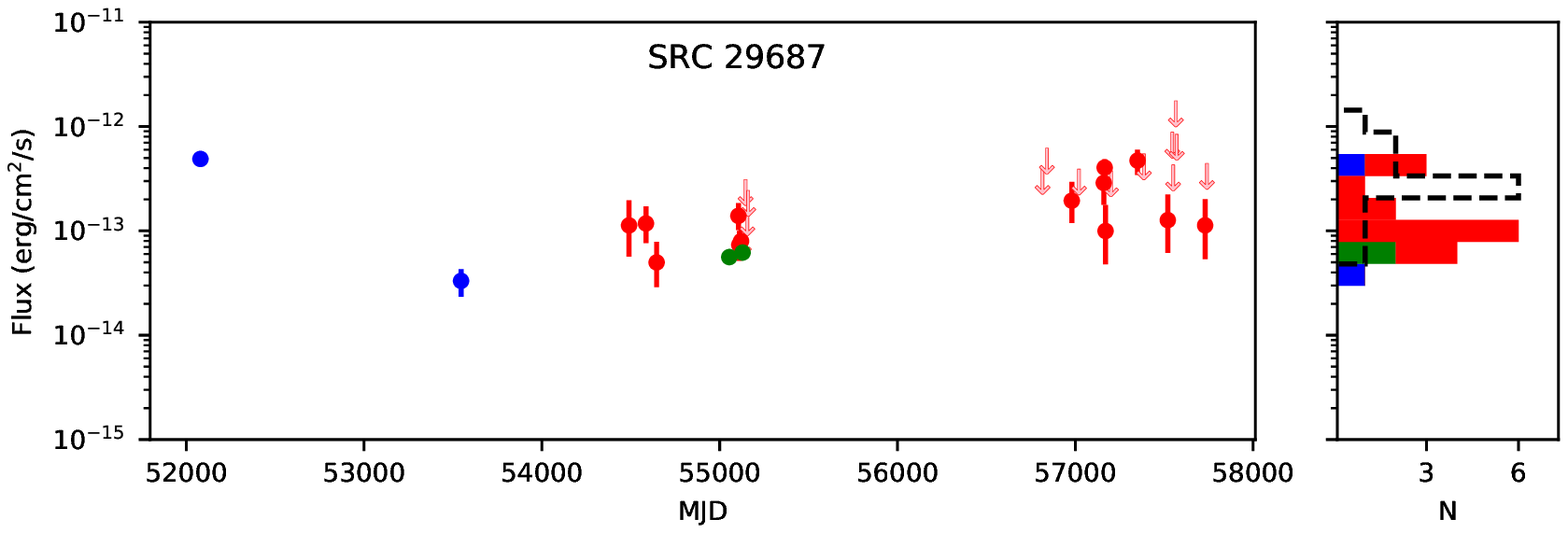}
    \includegraphics{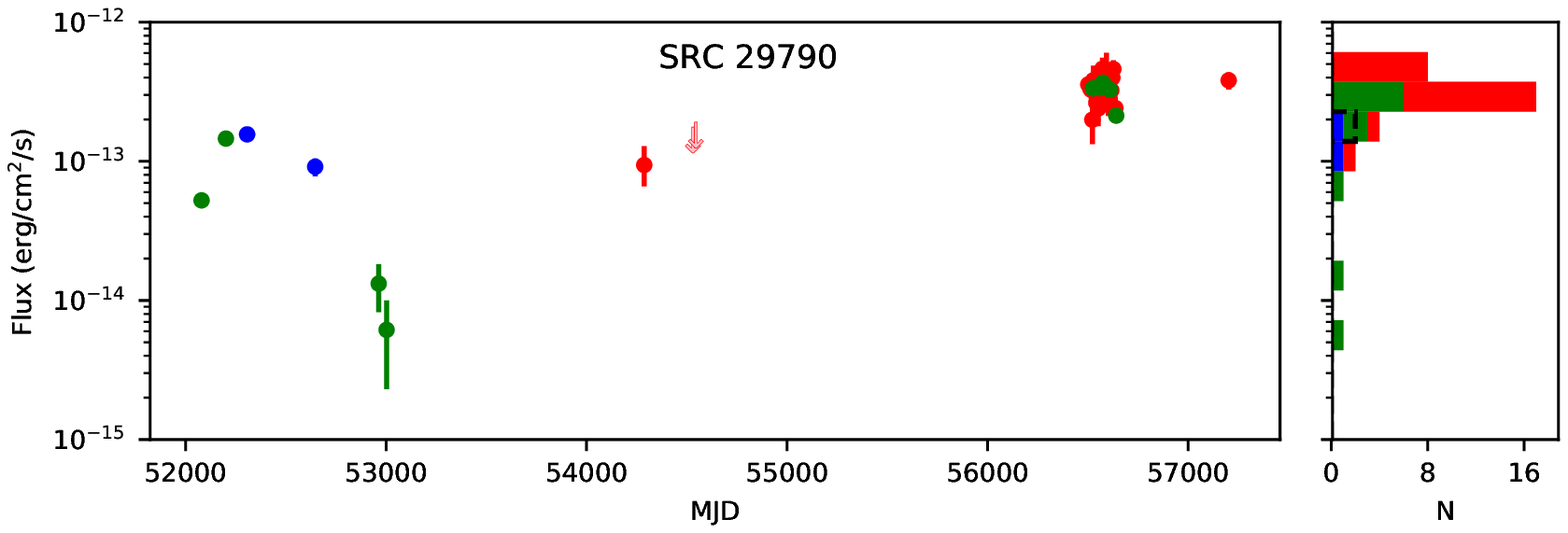}
    \includegraphics{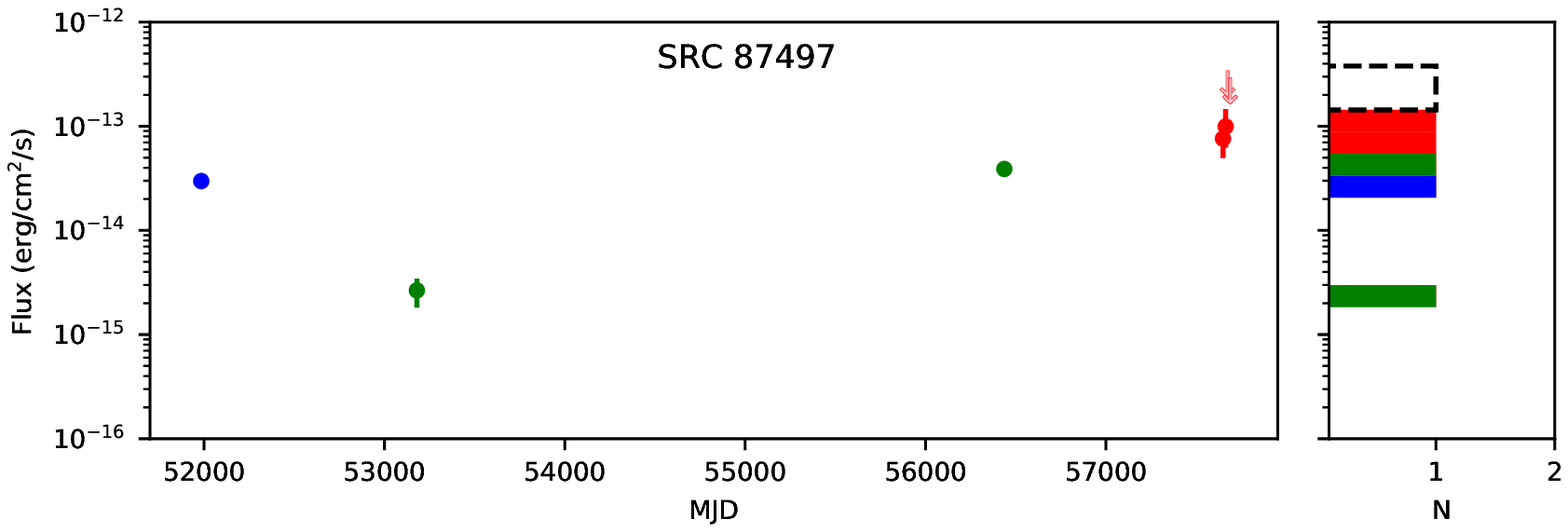}
    \includegraphics{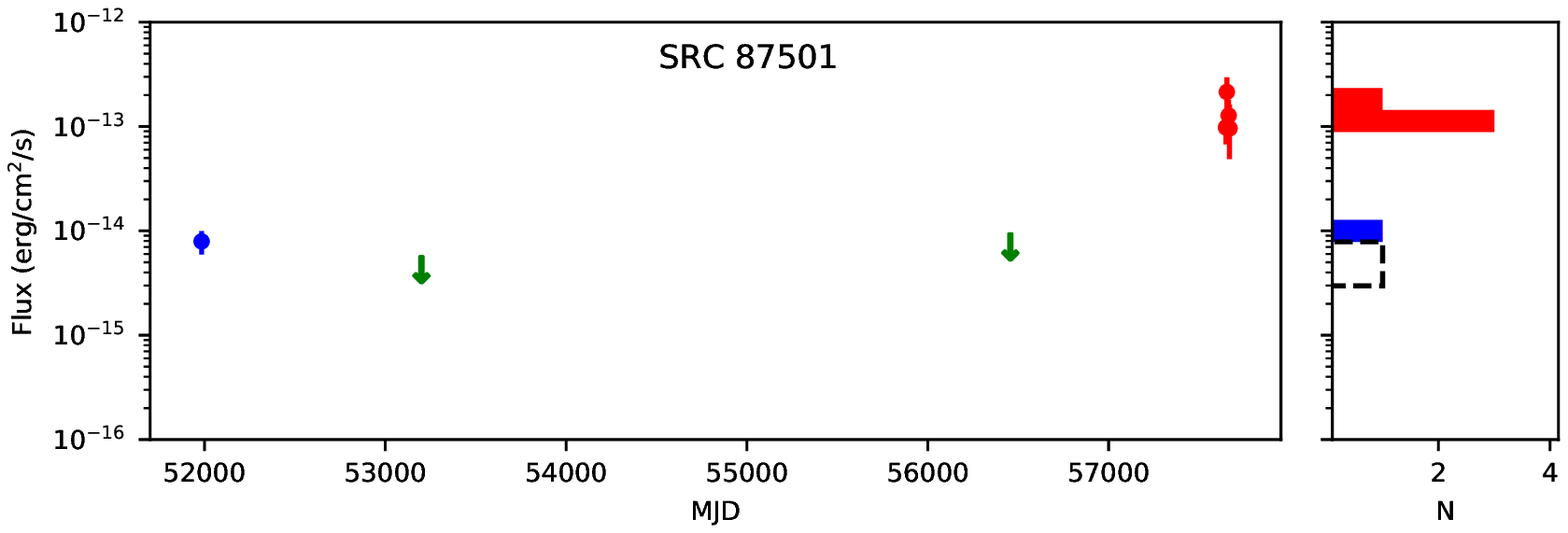}
\end{figure*}

\begin{figure*}
\centering
    \includegraphics{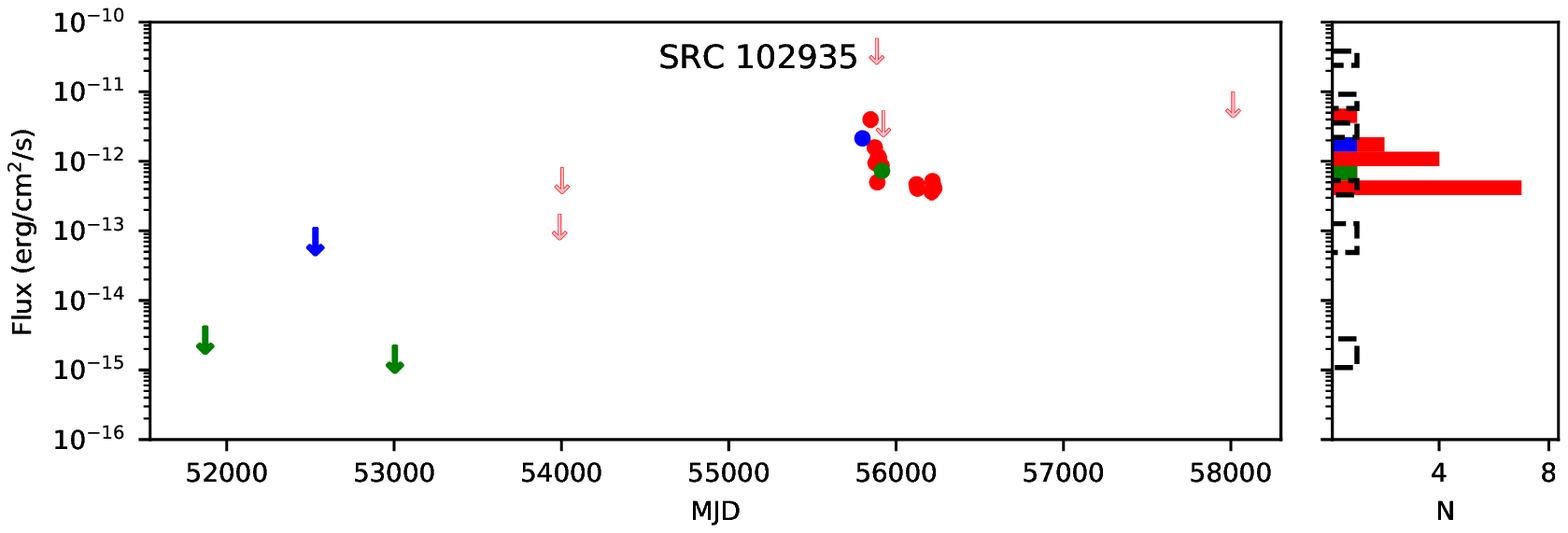}
    \includegraphics{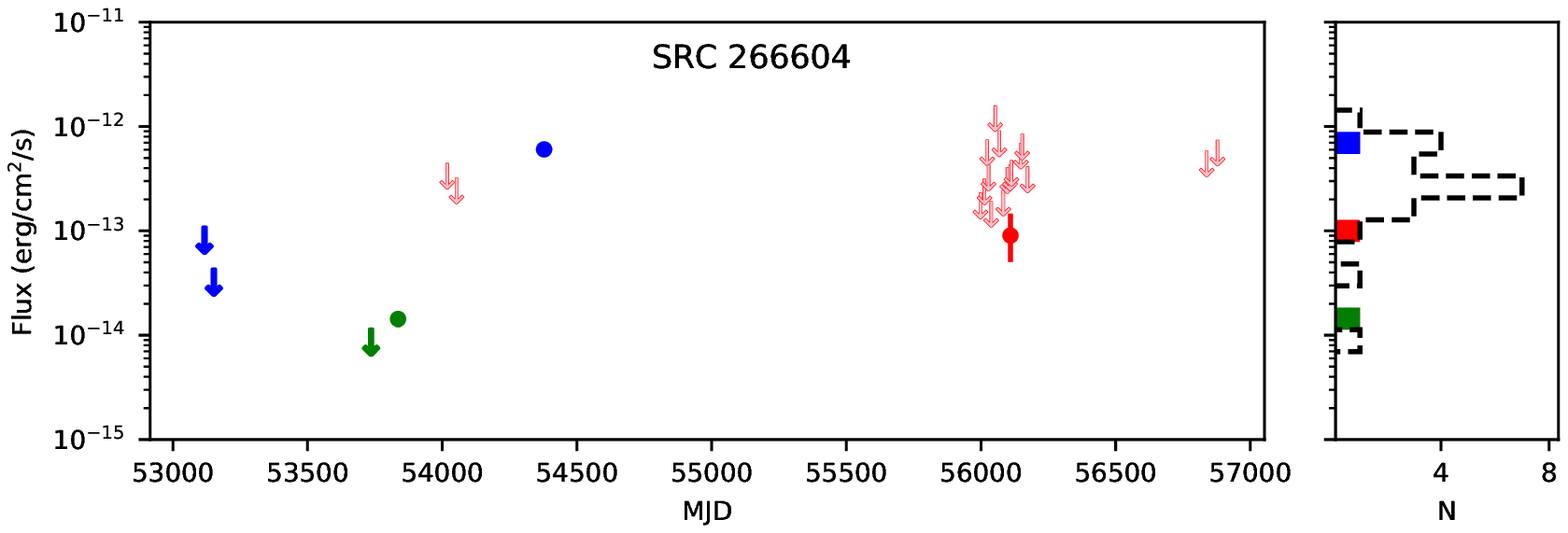}
    \includegraphics{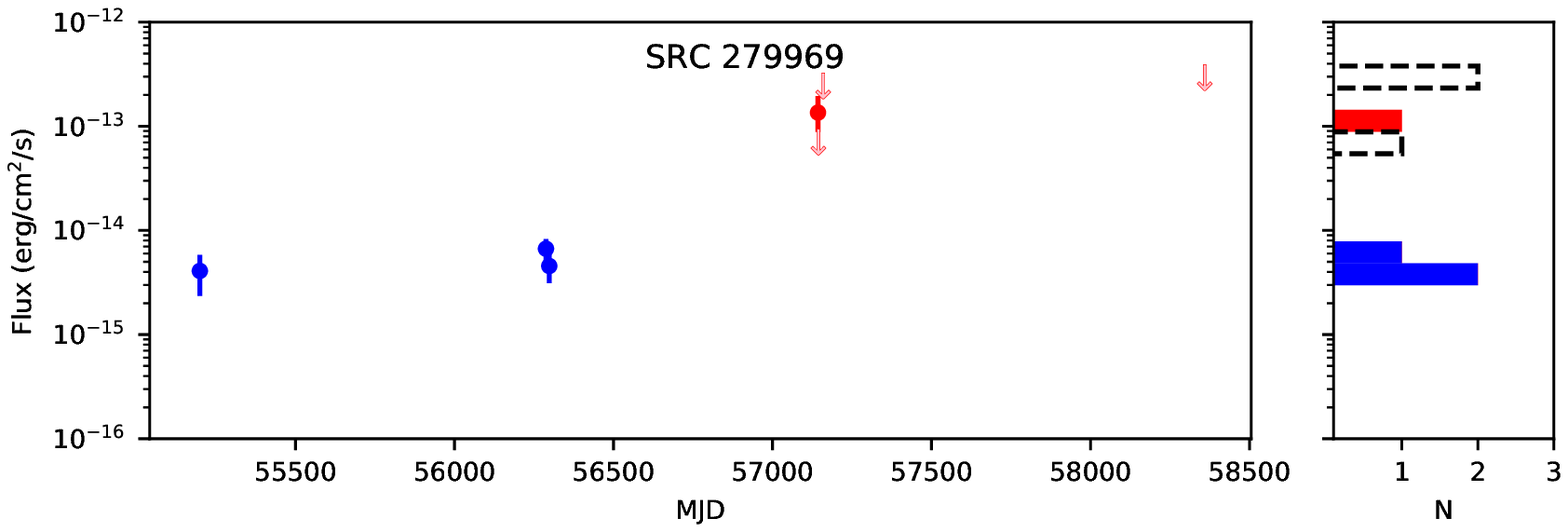}
    \includegraphics{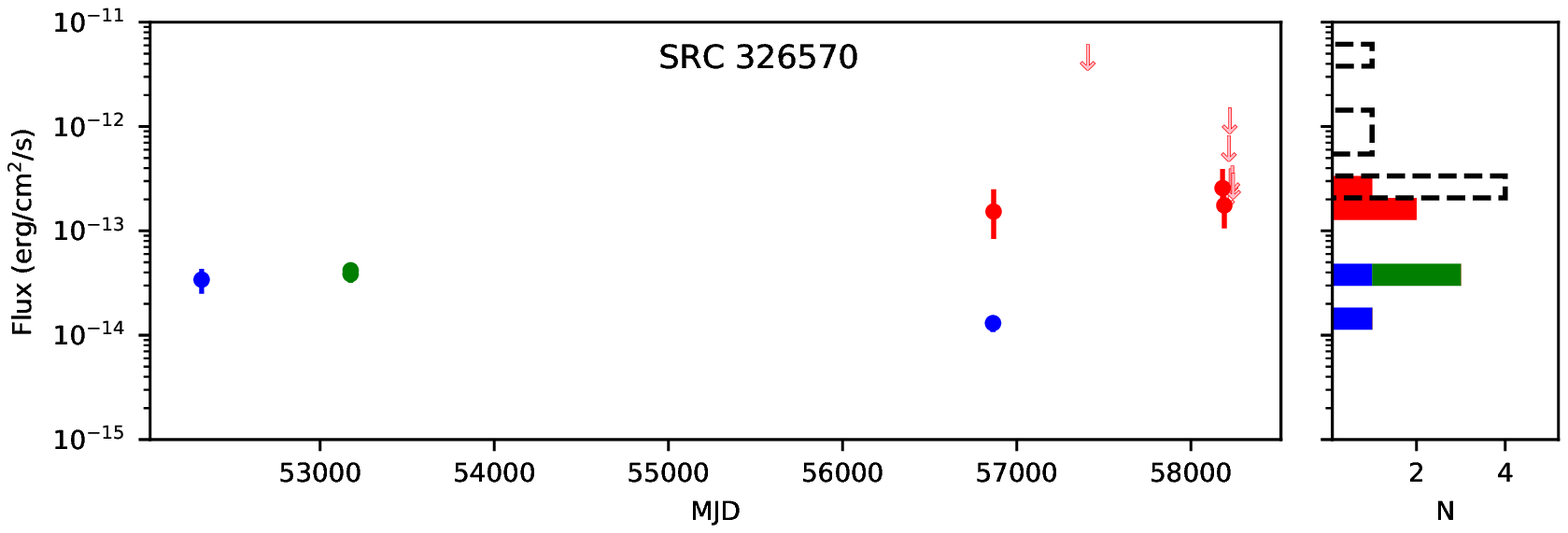}
\end{figure*}

\begin{figure*}
\centering
    \includegraphics{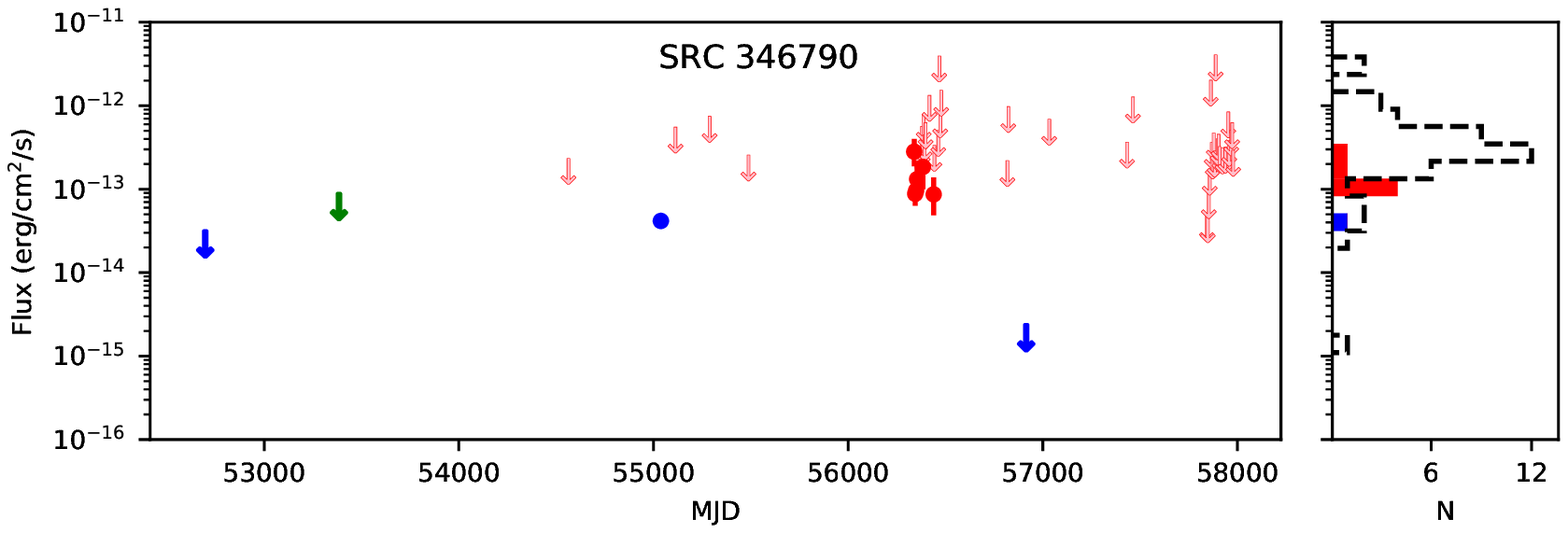}
    \includegraphics{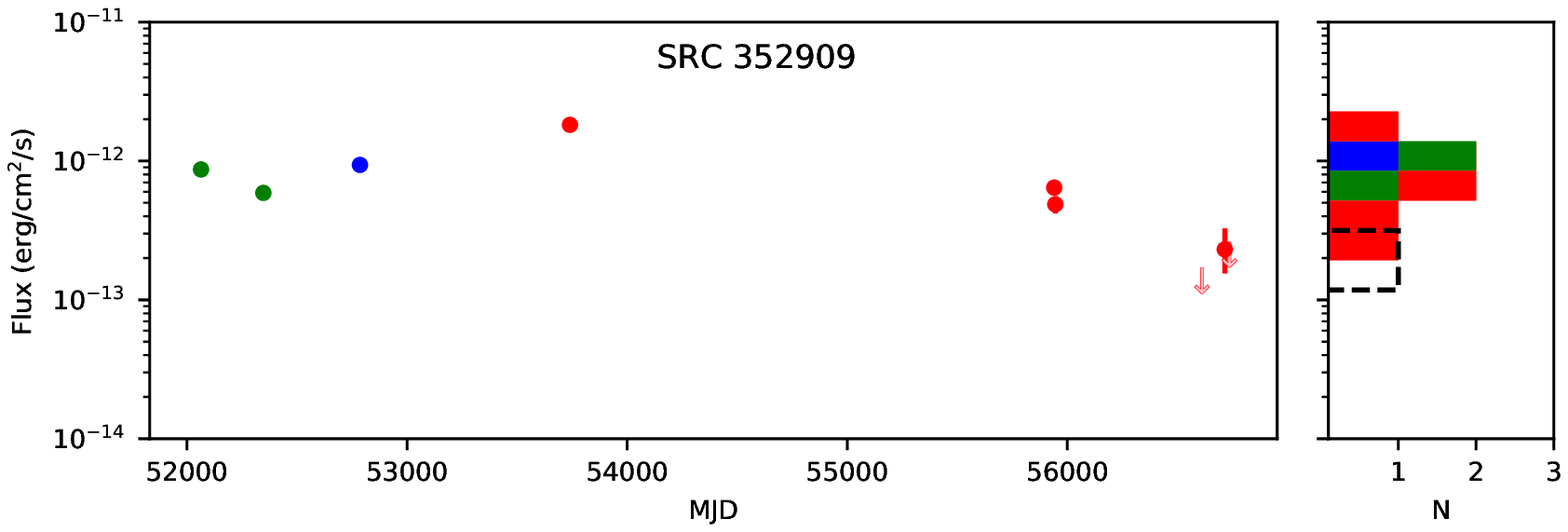}
    \includegraphics{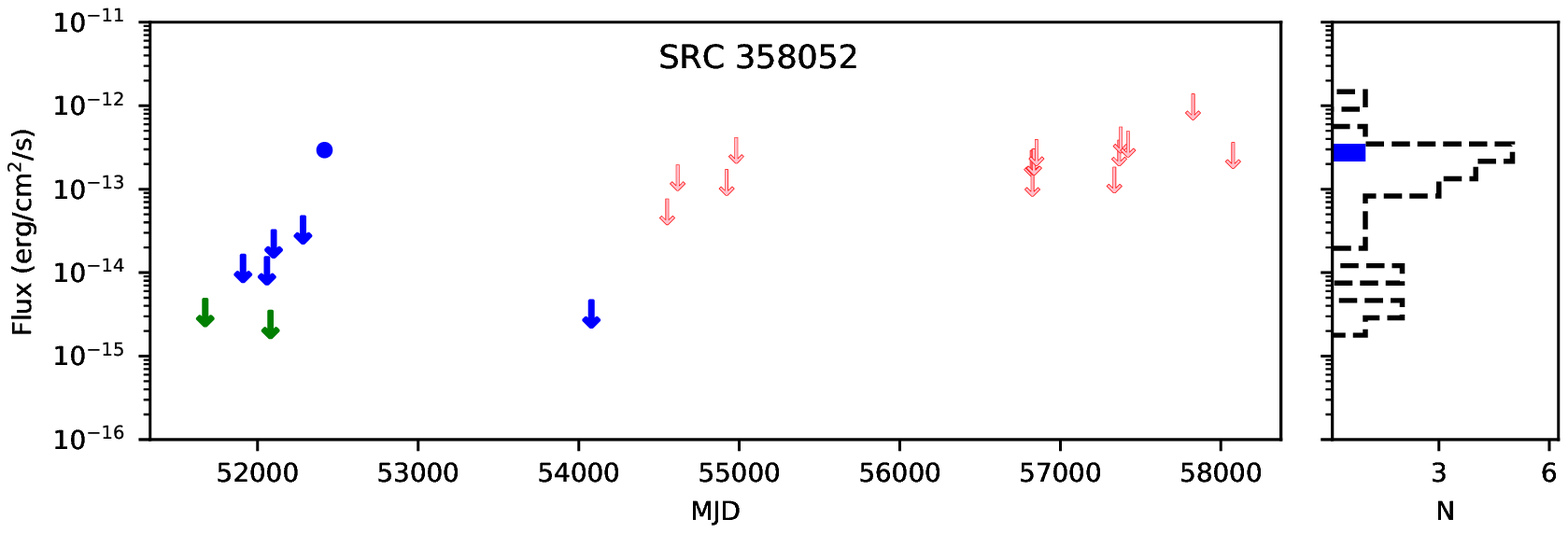}
    \includegraphics{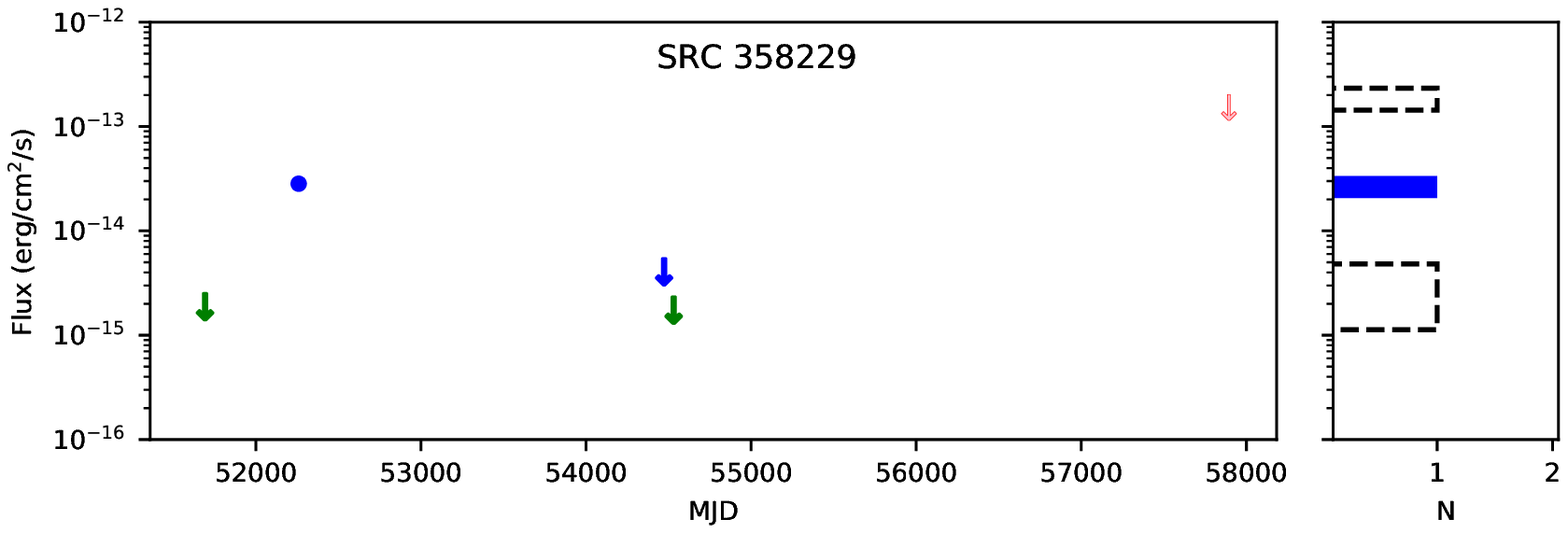}
\end{figure*}

\begin{figure*}
\centering
    \includegraphics{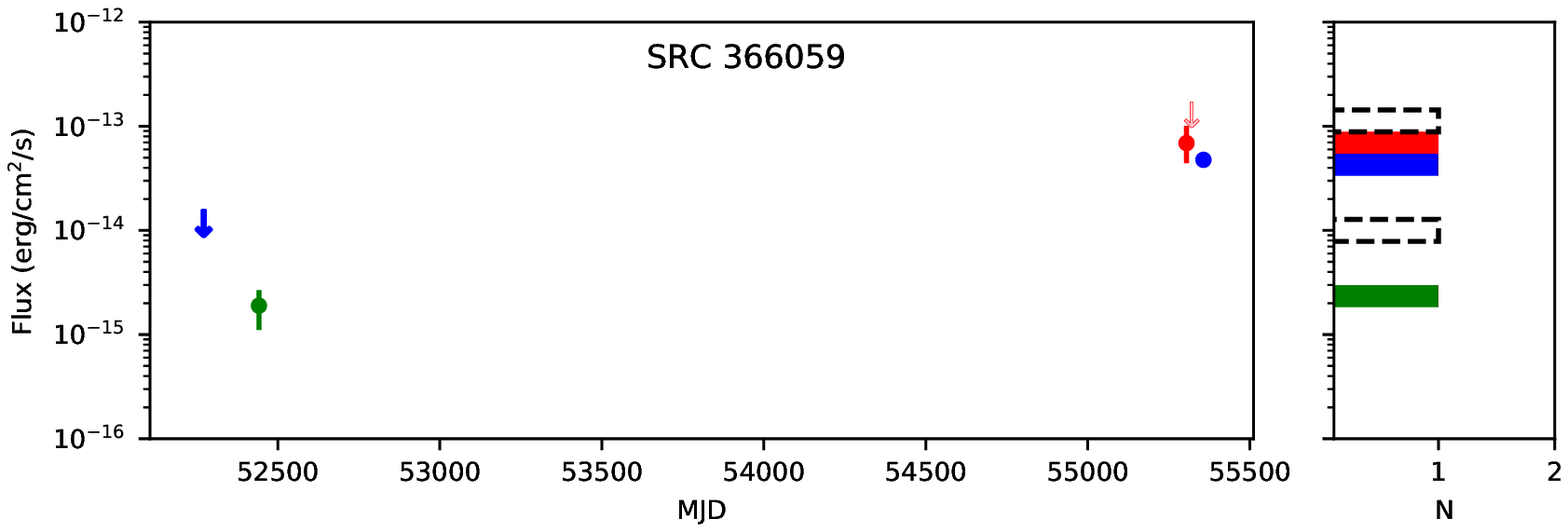}
\end{figure*}

\bibliographystyle{mnras}
\bibliography{ulx}

\begin{thebibliography}{}
\makeatletter
\relax
\def\mn@urlcharsother{\let\do\@makeother \do\$\do\&\do\#\do\^\do\_\do\%\do\~}
\def\mn@doi{\begingroup\mn@urlcharsother \@ifnextchar [ {\mn@doi@}
  {\mn@doi@[]}}
\def\mn@doi@[#1]#2{\def\@tempa{#1}\ifx\@tempa\@empty \href
  {http://dx.doi.org/#2} {doi:#2}\else \href {http://dx.doi.org/#2} {#1}\fi
  \endgroup}
\def\mn@eprint#1#2{\mn@eprint@#1:#2::\@nil}
\def\mn@eprint@arXiv#1{\href {http://arxiv.org/abs/#1} {{\tt arXiv:#1}}}
\def\mn@eprint@dblp#1{\href {http://dblp.uni-trier.de/rec/bibtex/#1.xml}
  {dblp:#1}}
\def\mn@eprint@#1:#2:#3:#4\@nil{\def\@tempa {#1}\def\@tempb {#2}\def\@tempc
  {#3}\ifx \@tempc \@empty \let \@tempc \@tempb \let \@tempb \@tempa \fi \ifx
  \@tempb \@empty \def\@tempb {arXiv}\fi \@ifundefined
  {mn@eprint@\@tempb}{\@tempb:\@tempc}{\expandafter \expandafter \csname
  mn@eprint@\@tempb\endcsname \expandafter{\@tempc}}}

\bibitem[\protect\citeauthoryear{{Bachetti} et~al.,}{{Bachetti}
  et~al.}{2014}]{Bachetti2014}
{Bachetti} M.,  et~al., 2014, \mn@doi [\nat] {10.1038/nature13791}, \href
  {http://adsabs.harvard.edu/abs/2014Natur.514..202B} {514, 202}

\bibitem[\protect\citeauthoryear{{Bachetti}, {Grefenstette}, {Walton},
  {Fuerst}, {Heida}, {Kennea}  \& {Lau}}{{Bachetti}
  et~al.}{2018}]{Bachetti2018}
{Bachetti} M.,  {Grefenstette} B.~W.,  {Walton} D.~J.,  {Fuerst} F.,  {Heida}
  M.,  {Kennea} J.~A.,   {Lau} R.,  2018, The Astronomer's Telegram, \href
  {http://adsabs.harvard.edu/abs/2018ATel11282....1B} {11282}

\bibitem[\protect\citeauthoryear{{Basko} \& {Sunyaev}}{{Basko} \&
  {Sunyaev}}{1976}]{Basko1976}
{Basko} M.~M.,  {Sunyaev} R.~A.,  1976, \mn@doi [\mnras]
  {10.1093/mnras/175.2.395}, \href
  {http://adsabs.harvard.edu/abs/1976MNRAS.175..395B} {175, 395}

\bibitem[\protect\citeauthoryear{{Bauer}, {Brandt}  \& {Lehmer}}{{Bauer}
  et~al.}{2003}]{Bauer2003}
{Bauer} F.~E.,  {Brandt} W.~N.,   {Lehmer} B.,  2003, \mn@doi [\aj]
  {10.1086/379139}, \href
  {https://ui.adsabs.harvard.edu/abs/2003AJ....126.2797B} {126, 2797}

\bibitem[\protect\citeauthoryear{{Brightman} et~al.,}{{Brightman}
  et~al.}{2016}]{Brightman2016}
{Brightman} M.,  et~al., 2016, \mn@doi [\apj] {10.3847/0004-637X/829/1/28},
  \href {http://adsabs.harvard.edu/abs/2016ApJ...829...28B} {829, 28}

\bibitem[\protect\citeauthoryear{{Brightman} et~al.,}{{Brightman}
  et~al.}{2018}]{Brightman2018}
{Brightman} M.,  et~al., 2018, \mn@doi [Nature Astronomy]
  {10.1038/s41550-018-0391-6}, \href
  {http://adsabs.harvard.edu/abs/2018NatAs...2..312B} {2, 312}

\bibitem[\protect\citeauthoryear{{Brightman} et~al.,}{{Brightman}
  et~al.}{2019}]{Brightman2019}
{Brightman} M.,  et~al., 2019, \mn@doi [\apj] {10.3847/1538-4357/ab0215}, \href
  {https://ui.adsabs.harvard.edu/abs/2019ApJ...873..115B} {873, 115}

\bibitem[\protect\citeauthoryear{{Burke} et~al.,}{{Burke}
  et~al.}{2013}]{Burke2013}
{Burke} M.~J.,  et~al., 2013, \mn@doi [\apj] {10.1088/0004-637X/775/1/21},
  \href {https://ui.adsabs.harvard.edu/abs/2013ApJ...775...21B} {775, 21}

\bibitem[\protect\citeauthoryear{{Burrows} et~al.,}{{Burrows}
  et~al.}{2005}]{Burrows2005}
{Burrows} D.~N.,  et~al., 2005, \mn@doi [\ssr] {10.1007/s11214-005-5097-2},
  \href {http://adsabs.harvard.edu/abs/2005SSRv..120..165B} {120, 165}

\bibitem[\protect\citeauthoryear{{Carpano}, {Haberl}, {Maitra}  \&
  {Vasilopoulos}}{{Carpano} et~al.}{2018}]{Carpano2018}
{Carpano} S.,  {Haberl} F.,  {Maitra} C.,   {Vasilopoulos} G.,  2018, \mn@doi
  [\mnras] {10.1093/mnrasl/sly030}, \href
  {http://adsabs.harvard.edu/abs/2018MNRAS.476L..45C} {476, L45}

\bibitem[\protect\citeauthoryear{{Carrera} et~al.,}{{Carrera}
  et~al.}{2007}]{Carrera2007}
{Carrera} F.~J.,  et~al., 2007, \mn@doi [\aap] {10.1051/0004-6361:20066271},
  \href {http://adsabs.harvard.edu/abs/2007A%26A...469...27C} {469, 27}

\bibitem[\protect\citeauthoryear{{Dall'Osso}, {Perna}  \& {Stella}}{{Dall'Osso}
  et~al.}{2015}]{DallOsso2015}
{Dall'Osso} S.,  {Perna} R.,   {Stella} L.,  2015, \mn@doi [\mnras]
  {10.1093/mnras/stv170}, \href
  {http://adsabs.harvard.edu/abs/2015MNRAS.449.2144D} {449, 2144}

\bibitem[\protect\citeauthoryear{{Earnshaw}, {Roberts}  \&
  {Sathyaprakash}}{{Earnshaw} et~al.}{2018}]{Earnshaw2018}
{Earnshaw} H.~P.,  {Roberts} T.~P.,   {Sathyaprakash} R.,  2018, \mn@doi
  [\mnras] {10.1093/mnras/sty501}, \href
  {http://adsabs.harvard.edu/abs/2018MNRAS.tmp..485E} {}

\bibitem[\protect\citeauthoryear{{Earnshaw}, {Roberts}, {Middleton}, {Walton}
  \& {Mateos}}{{Earnshaw} et~al.}{2019}]{Earnshaw19}
{Earnshaw} H.~P.,  {Roberts} T.~P.,  {Middleton} M.~J.,  {Walton} D.~J.,
  {Mateos} S.,  2019, \mn@doi [\mnras] {10.1093/mnras/sty3403}, \href
  {http://adsabs.harvard.edu/abs/2019MNRAS.483.5554E} {483, 5554}

\bibitem[\protect\citeauthoryear{{Ek{\c s}i}, {Anda{\c c}}, {{\c
  C}{\i}k{\i}nto{\u g}lu}, {Gen{\c c}ali}, {G{\"u}ng{\"o}r}  \&
  {{\"O}ztekin}}{{Ek{\c s}i} et~al.}{2015}]{Eksi2015}
{Ek{\c s}i} K.~Y.,  {Anda{\c c}} {\.I}.~C.,  {{\c C}{\i}k{\i}nto{\u g}lu} S.,
  {Gen{\c c}ali} A.~A.,  {G{\"u}ng{\"o}r} C.,   {{\"O}ztekin} F.,  2015,
  \mn@doi [\mnras] {10.1093/mnrasl/slu199}, \href
  {http://adsabs.harvard.edu/abs/2015MNRAS.448L..40E} {448, L40}

\bibitem[\protect\citeauthoryear{{Evans} et~al.,}{{Evans}
  et~al.}{2007}]{Evans2007}
{Evans} P.~A.,  et~al., 2007, \mn@doi [\aap] {10.1051/0004-6361:20077530},
  \href {https://ui.adsabs.harvard.edu/abs/2007A&A...469..379E} {469, 379}

\bibitem[\protect\citeauthoryear{{Evans} et~al.,}{{Evans}
  et~al.}{2009}]{Evans2009}
{Evans} P.~A.,  et~al., 2009, \mn@doi [\mnras]
  {10.1111/j.1365-2966.2009.14913.x}, \href
  {http://adsabs.harvard.edu/abs/2009MNRAS.397.1177E} {397, 1177}

\bibitem[\protect\citeauthoryear{{F{\"u}rst} et~al.,}{{F{\"u}rst}
  et~al.}{2016}]{Furst2016}
{F{\"u}rst} F.,  et~al., 2016, \mn@doi [\apjl] {10.3847/2041-8205/831/2/L14},
  \href {http://adsabs.harvard.edu/abs/2016ApJ...831L..14F} {831, L14}

\bibitem[\protect\citeauthoryear{{F{\"u}rst}, {Walton}, {Stern}, {Bachetti},
  {Barret}, {Brightman}, {Harrison}  \& {Rana}}{{F{\"u}rst}
  et~al.}{2017}]{Furst2017}
{F{\"u}rst} F.,  {Walton} D.~J.,  {Stern} D.,  {Bachetti} M.,  {Barret} D.,
  {Brightman} M.,  {Harrison} F.~A.,   {Rana} V.,  2017, \mn@doi [\apj]
  {10.3847/1538-4357/834/1/77}, \href
  {http://adsabs.harvard.edu/abs/2017ApJ...834...77F} {834, 77}

\bibitem[\protect\citeauthoryear{{F{\"u}rst} et~al.,}{{F{\"u}rst}
  et~al.}{2018}]{Fuerst18}
{F{\"u}rst} F.,  et~al., 2018, \mn@doi [\aap] {10.1051/0004-6361/201833292},
  \href {http://adsabs.harvard.edu/abs/2018A%26A...616A.186F} {616, A186}

\bibitem[\protect\citeauthoryear{{Garmire}, {Bautz}, {Ford}, {Nousek}  \&
  {Ricker}}{{Garmire} et~al.}{2003}]{CHANDRA_ACIS}
{Garmire} G.~P.,  {Bautz} M.~W.,  {Ford} P.~G.,  {Nousek} J.~A.,   {Ricker} Jr.
  G.~R.,  2003, in {Truemper} J.~E.,  {Tananbaum} H.~D.,  eds,  SPIE Conference
  Series Vol. 4851, SPIE Conference Series. pp 28--44,
  \mn@doi{10.1117/12.461599}

\bibitem[\protect\citeauthoryear{{Gehrels} et~al.,}{{Gehrels}
  et~al.}{2004}]{SWIFT}
{Gehrels} N.,  et~al., 2004, \mn@doi [\apj] {10.1086/422091}, \href
  {http://adsabs.harvard.edu/abs/2004ApJ...611.1005G} {611, 1005}

\bibitem[\protect\citeauthoryear{{Ghosh}, {Lamb}  \& {Pethick}}{{Ghosh}
  et~al.}{1977}]{Ghosh1977}
{Ghosh} P.,  {Lamb} F.~K.,   {Pethick} C.~J.,  1977, \mn@doi [\apj]
  {10.1086/155606}, \href {http://adsabs.harvard.edu/abs/1977ApJ...217..578G}
  {217, 578}

\bibitem[\protect\citeauthoryear{{Gladstone}, {Roberts}  \& {Done}}{{Gladstone}
  et~al.}{2009}]{Gladstone2009}
{Gladstone} J.~C.,  {Roberts} T.~P.,   {Done} C.,  2009, \mn@doi [\mnras]
  {10.1111/j.1365-2966.2009.15123.x}, \href
  {https://ui.adsabs.harvard.edu/abs/2009MNRAS.397.1836G} {397, 1836}

\bibitem[\protect\citeauthoryear{{Herold}}{{Herold}}{1979}]{Herold1979}
{Herold} H.,  1979, \mn@doi [\prd] {10.1103/PhysRevD.19.2868}, \href
  {http://adsabs.harvard.edu/abs/1979PhRvD..19.2868H} {19, 2868}

\bibitem[\protect\citeauthoryear{{Hodges-Kluck}, {Bregman}, {Miller}  \&
  {Pellegrini}}{{Hodges-Kluck} et~al.}{2012}]{HodgesKluck12}
{Hodges-Kluck} E.~J.,  {Bregman} J.~N.,  {Miller} J.~M.,   {Pellegrini} E.,
  2012, \mn@doi [\apjl] {10.1088/2041-8205/747/2/L39}, \href
  {http://adsabs.harvard.edu/abs/2012ApJ...747L..39H} {747, L39}

\bibitem[\protect\citeauthoryear{{Israel} et~al.,}{{Israel}
  et~al.}{2017a}]{IsraelBelfiore2017}
{Israel} G.~L.,  et~al., 2017a, \mn@doi [Science] {10.1126/science.aai8635},
  \href {http://adsabs.harvard.edu/abs/2017Sci...355..817I} {355, 817}

\bibitem[\protect\citeauthoryear{{Israel} et~al.,}{{Israel}
  et~al.}{2017b}]{Israel2017}
{Israel} G.~L.,  et~al., 2017b, \mn@doi [\mnras] {10.1093/mnrasl/slw218}, \href
  {http://adsabs.harvard.edu/abs/2017MNRAS.466L..48I} {466, L48}

\bibitem[\protect\citeauthoryear{{Jansen} et~al.,}{{Jansen} et~al.}{2001}]{XMM}
{Jansen} F.,  et~al., 2001, \mn@doi [\aap] {10.1051/0004-6361:20000036}, \href
  {http://adsabs.harvard.edu/abs/2001A%26A...365L...1J} {365, L1}

\bibitem[\protect\citeauthoryear{{Kaaret}, {Feng}  \& {Roberts}}{{Kaaret}
  et~al.}{2017}]{Kaaret2017}
{Kaaret} P.,  {Feng} H.,   {Roberts} T.~P.,  2017, \mn@doi [\araa]
  {10.1146/annurev-astro-091916-055259}, \href
  {http://adsabs.harvard.edu/abs/2017ARA%26A..55..303K} {55, 303}

\bibitem[\protect\citeauthoryear{{Karachentsev}, {Karachentseva}, {Huchtmeier}
  \& {Makarov}}{{Karachentsev} et~al.}{2004}]{Karachentsev2004}
{Karachentsev} I.~D.,  {Karachentseva} V.~E.,  {Huchtmeier} W.~K.,   {Makarov}
  D.~I.,  2004, \mn@doi [\aj] {10.1086/382905}, \href
  {http://adsabs.harvard.edu/abs/2004AJ....127.2031K} {127, 2031}

\bibitem[\protect\citeauthoryear{{King} \& {Lasota}}{{King} \&
  {Lasota}}{2016}]{King2016}
{King} A.,  {Lasota} J.-P.,  2016, \mn@doi [\mnras] {10.1093/mnrasl/slw011},
  \href {http://adsabs.harvard.edu/abs/2016MNRAS.458L..10K} {458, L10}

\bibitem[\protect\citeauthoryear{{Klu{\'z}niak} \& {Lasota}}{{Klu{\'z}niak} \&
  {Lasota}}{2015}]{Kluzniak2015}
{Klu{\'z}niak} W.,  {Lasota} J.-P.,  2015, \mn@doi [\mnras]
  {10.1093/mnrasl/slu200}, \href
  {http://adsabs.harvard.edu/abs/2015MNRAS.448L..43K} {448, L43}

\bibitem[\protect\citeauthoryear{{Koliopanos}, {Vasilopoulos}, {Godet},
  {Bachetti}, {Webb}  \& {Barret}}{{Koliopanos} et~al.}{2017}]{Koliopanos17}
{Koliopanos} F.,  {Vasilopoulos} G.,  {Godet} O.,  {Bachetti} M.,  {Webb}
  N.~A.,   {Barret} D.,  2017, \mn@doi [\aap] {10.1051/0004-6361/201730922},
  \href {http://adsabs.harvard.edu/abs/2017A%26A...608A..47K} {608, A47}

\bibitem[\protect\citeauthoryear{{Koliopanos}, {Vasilopoulos}, {Buchner},
  {Maitra}  \& {Haberl}}{{Koliopanos} et~al.}{2019}]{Koliopanos2019}
{Koliopanos} F.,  {Vasilopoulos} G.,  {Buchner} J.,  {Maitra} C.,   {Haberl}
  F.,  2019, \mn@doi [\aap] {10.1051/0004-6361/201834144}, \href
  {https://ui.adsabs.harvard.edu/abs/2019A&A...621A.118K} {621, A118}

\bibitem[\protect\citeauthoryear{{Kotze} \& {Charles}}{{Kotze} \&
  {Charles}}{2012}]{Kotze12}
{Kotze} M.~M.,  {Charles} P.~A.,  2012, \mn@doi [\mnras]
  {10.1111/j.1365-2966.2011.20146.x}, \href
  {http://adsabs.harvard.edu/abs/2012MNRAS.420.1575K} {420, 1575}

\bibitem[\protect\citeauthoryear{{Kraft}, {Burrows}  \& {Nousek}}{{Kraft}
  et~al.}{1991}]{Kraft1991}
{Kraft} R.~P.,  {Burrows} D.~N.,   {Nousek} J.~A.,  1991, \mn@doi [\apj]
  {10.1086/170124}, \href {http://adsabs.harvard.edu/abs/1991ApJ...374..344K}
  {374, 344}

\bibitem[\protect\citeauthoryear{{Liu} \& {Bregman}}{{Liu} \&
  {Bregman}}{2005}]{LiuB2005}
{Liu} J.-F.,  {Bregman} J.~N.,  2005, \mn@doi [\apjs] {10.1086/427170}, \href
  {https://ui.adsabs.harvard.edu/abs/2005ApJS..157...59L} {157, 59}

\bibitem[\protect\citeauthoryear{{Liu} \& {Mirabel}}{{Liu} \&
  {Mirabel}}{2005}]{LiuM2005}
{Liu} Q.~Z.,  {Mirabel} I.~F.,  2005, \mn@doi [\aap]
  {10.1051/0004-6361:20041878}, \href
  {https://ui.adsabs.harvard.edu/abs/2005A&A...429.1125L} {429, 1125}

\bibitem[\protect\citeauthoryear{{Madsen} et~al.,}{{Madsen}
  et~al.}{2015}]{Madsen15}
{Madsen} K.~K.,  et~al., 2015, \mn@doi [\apjs] {10.1088/0067-0049/220/1/8},
  \href {http://adsabs.harvard.edu/abs/2015ApJS..220....8M} {220, 8}

\bibitem[\protect\citeauthoryear{{Madsen}, {Beardmore}, {Forster}, {Guainazzi},
  {Marshall}, {Miller}, {Page}  \& {Stuhlinger}}{{Madsen}
  et~al.}{2017}]{Madsen17}
{Madsen} K.~K.,  {Beardmore} A.~P.,  {Forster} K.,  {Guainazzi} M.,  {Marshall}
  H.~L.,  {Miller} E.~D.,  {Page} K.~L.,   {Stuhlinger} M.,  2017, \mn@doi
  [\aj] {10.3847/1538-3881/153/1/2}, \href
  {http://adsabs.harvard.edu/abs/2017AJ....153....2M} {153, 2}

\bibitem[\protect\citeauthoryear{{Merloni} et~al.,}{{Merloni}
  et~al.}{2012}]{EROSITA_tmp}
{Merloni} A.,  et~al., 2012, preprint, \href
  {http://adsabs.harvard.edu/abs/2012arXiv1209.3114M} {} (\mn@eprint {arXiv}
  {1209.3114})

\bibitem[\protect\citeauthoryear{{Middleton} \& {King}}{{Middleton} \&
  {King}}{2017}]{Middleton17}
{Middleton} M.~J.,  {King} A.,  2017, \mn@doi [\mnras] {10.1093/mnrasl/slx079},
  \href {http://adsabs.harvard.edu/abs/2017MNRAS.470L..69M} {470, L69}

\bibitem[\protect\citeauthoryear{{Middleton} et~al.,}{{Middleton}
  et~al.}{2013}]{Middleton2013}
{Middleton} M.~J.,  et~al., 2013, \mn@doi [\nat] {10.1038/nature11697}, \href
  {https://ui.adsabs.harvard.edu/abs/2013Natur.493..187M} {493, 187}

\bibitem[\protect\citeauthoryear{{Middleton}, {Brightman}, {Pintore},
  {Bachetti}, {Fabian}, {F{\"u}rst}  \& {Walton}}{{Middleton}
  et~al.}{2019}]{Middleton19}
{Middleton} M.~J.,  {Brightman} M.,  {Pintore} F.,  {Bachetti} M.,  {Fabian}
  A.~C.,  {F{\"u}rst} F.,   {Walton} D.~J.,  2019, \mn@doi [\mnras]
  {10.1093/mnras/stz436}, \href
  {http://adsabs.harvard.edu/abs/2019MNRAS.tmp..468M} {}

\bibitem[\protect\citeauthoryear{{Monard}}{{Monard}}{2010}]{Monard2010}
{Monard} L.~A.~G.,  2010, Central Bureau Electronic Telegrams, \href
  {http://adsabs.harvard.edu/abs/2010CBET.2289....1M} {2289}

\bibitem[\protect\citeauthoryear{{Motch}, {Pakull}, {Soria}, {Gris{\'e}}  \&
  {Pietrzy{\'n}ski}}{{Motch} et~al.}{2014}]{Motch2014}
{Motch} C.,  {Pakull} M.~W.,  {Soria} R.,  {Gris{\'e}} F.,   {Pietrzy{\'n}ski}
  G.,  2014, \mn@doi [\nat] {10.1038/nature13730}, \href
  {http://adsabs.harvard.edu/abs/2014Natur.514..198M} {514, 198}

\bibitem[\protect\citeauthoryear{{Mukai}}{{Mukai}}{1993}]{PIMMS}
{Mukai} K.,  1993, Legacy, vol.~3, p.21-31, \href
  {http://adsabs.harvard.edu/abs/1993Legac...3...21M} {3, 21}

\bibitem[\protect\citeauthoryear{{Mushtukov}, {Suleimanov}, {Tsygankov}  \&
  {Poutanen}}{{Mushtukov} et~al.}{2015}]{Mushtukov2015}
{Mushtukov} A.~A.,  {Suleimanov} V.~F.,  {Tsygankov} S.~S.,   {Poutanen} J.,
  2015, \mn@doi [\mnras] {10.1093/mnras/stv2087}, \href
  {http://adsabs.harvard.edu/abs/2015MNRAS.454.2539M} {454, 2539}

\bibitem[\protect\citeauthoryear{{Nandra} et~al.,}{{Nandra}
  et~al.}{2013}]{ATHENA}
{Nandra} K.,  et~al., 2013, preprint, \href
  {http://adsabs.harvard.edu/abs/2013arXiv1306.2307N} {} (\mn@eprint {arXiv}
  {1306.2307})

\bibitem[\protect\citeauthoryear{{Paolillo} et~al.,}{{Paolillo}
  et~al.}{2017}]{Paolillo2017}
{Paolillo} M.,  et~al., 2017, \mn@doi [\mnras] {10.1093/mnras/stx1761}, \href
  {https://ui.adsabs.harvard.edu/abs/2017MNRAS.471.4398P} {471, 4398}

\bibitem[\protect\citeauthoryear{{Pintore}, {Zampieri}, {Stella}, {Wolter},
  {Mereghetti}  \& {Israel}}{{Pintore} et~al.}{2017}]{Pintore2017}
{Pintore} F.,  {Zampieri} L.,  {Stella} L.,  {Wolter} A.,  {Mereghetti} S.,
  {Israel} G.~L.,  2017, \mn@doi [\apj] {10.3847/1538-4357/836/1/113}, \href
  {http://adsabs.harvard.edu/abs/2017ApJ...836..113P} {836, 113}

\bibitem[\protect\citeauthoryear{{Plucinsky}, {Bogdan}, {Marshall}  \&
  {Tice}}{{Plucinsky} et~al.}{2018}]{Plucinsky18}
{Plucinsky} P.~P.,  {Bogdan} A.,  {Marshall} H.~L.,   {Tice} N.~W.,  2018, in
  Space Telescopes and Instrumentation 2018: Ultraviolet to Gamma Ray. p.
  106996B (\mn@eprint {arXiv} {1809.02225}), \mn@doi{10.1117/12.2312748}

\bibitem[\protect\citeauthoryear{{Rodr{\'\i}guez Castillo}
  et~al.,}{{Rodr{\'\i}guez Castillo} et~al.}{2019}]{Rodriguez19}
{Rodr{\'\i}guez Castillo} G.~A.,  et~al., 2019, arXiv e-prints, \href
  {https://ui.adsabs.harvard.edu/abs/2019arXiv190604791R} {p. arXiv:1906.04791}

\bibitem[\protect\citeauthoryear{{Rosen} et~al.,}{{Rosen}
  et~al.}{2016}]{Rosen2016}
{Rosen} S.~R.,  et~al., 2016, \mn@doi [\aap] {10.1051/0004-6361/201526416},
  \href {http://adsabs.harvard.edu/abs/2016A%26A...590A...1R} {590, A1}

\bibitem[\protect\citeauthoryear{{Sathyaprakash} et~al.,}{{Sathyaprakash}
  et~al.}{2019}]{Sathyaprakash19}
{Sathyaprakash} R.,  et~al., 2019, arXiv e-prints, \href
  {https://ui.adsabs.harvard.edu/abs/2019arXiv190600640S} {p. arXiv:1906.00640}

\bibitem[\protect\citeauthoryear{{Strateva} \& {Komossa}}{{Strateva} \&
  {Komossa}}{2009}]{Strateva2009}
{Strateva} I.~V.,  {Komossa} S.,  2009, \mn@doi [\apj]
  {10.1088/0004-637X/692/1/443}, \href
  {https://ui.adsabs.harvard.edu/abs/2009ApJ...692..443S} {692, 443}

\bibitem[\protect\citeauthoryear{{Str{\"u}der} et~al.,}{{Str{\"u}der}
  et~al.}{2001}]{XMM_PN}
{Str{\"u}der} L.,  et~al., 2001, \mn@doi [\aap] {10.1051/0004-6361:20000066},
  \href {http://adsabs.harvard.edu/abs/2001A%26A...365L..18S} {365, L18}

\bibitem[\protect\citeauthoryear{{Tsygankov}, {Mushtukov}, {Suleimanov}  \&
  {Poutanen}}{{Tsygankov} et~al.}{2016}]{Tsygankov2016}
{Tsygankov} S.~S.,  {Mushtukov} A.~A.,  {Suleimanov} V.~F.,   {Poutanen} J.,
  2016, \mn@doi [\mnras] {10.1093/mnras/stw046}, \href
  {http://adsabs.harvard.edu/abs/2016MNRAS.457.1101T} {457, 1101}

\bibitem[\protect\citeauthoryear{{Turner} et~al.,}{{Turner}
  et~al.}{2001}]{XMM_MOS}
{Turner} M.~J.~L.,  et~al., 2001, \mn@doi [\aap] {10.1051/0004-6361:20000087},
  \href {http://adsabs.harvard.edu/abs/2001A%26A...365L..27T} {365, L27}

\bibitem[\protect\citeauthoryear{{Uttley}, {McHardy}  \& {Vaughan}}{{Uttley}
  et~al.}{2005}]{Uttley05}
{Uttley} P.,  {McHardy} I.~M.,   {Vaughan} S.,  2005, \mn@doi [\mnras]
  {10.1111/j.1365-2966.2005.08886.x}, \href
  {http://adsabs.harvard.edu/abs/2005MNRAS.359..345U} {359, 345}

\bibitem[\protect\citeauthoryear{{Walton}, {Roberts}, {Mateos}  \&
  {Heard}}{{Walton} et~al.}{2011}]{Walton2011}
{Walton} D.~J.,  {Roberts} T.~P.,  {Mateos} S.,   {Heard} V.,  2011, \mn@doi
  [\mnras] {10.1111/j.1365-2966.2011.19154.x}, \href
  {http://adsabs.harvard.edu/abs/2011MNRAS.416.1844W} {416, 1844}

\bibitem[\protect\citeauthoryear{{Walton} et~al.,}{{Walton}
  et~al.}{2015}]{Walton2015}
{Walton} D.~J.,  et~al., 2015, \mn@doi [\apj] {10.1088/0004-637X/799/2/122},
  \href {http://adsabs.harvard.edu/abs/2015ApJ...799..122W} {799, 122}

\bibitem[\protect\citeauthoryear{{Walton} et~al.,}{{Walton}
  et~al.}{2016}]{Walton2016}
{Walton} D.~J.,  et~al., 2016, \mn@doi [\apjl] {10.3847/2041-8205/827/1/L13},
  \href {http://adsabs.harvard.edu/abs/2016ApJ...827L..13W} {827, L13}

\bibitem[\protect\citeauthoryear{{Walton} et~al.,}{{Walton}
  et~al.}{2018a}]{Walton2018}
{Walton} D.~J.,  et~al., 2018a, \mn@doi [\apj] {10.3847/1538-4357/aab610},
  \href {http://adsabs.harvard.edu/abs/2018ApJ...856..128W} {856, 128}

\bibitem[\protect\citeauthoryear{{Walton} et~al.,}{{Walton}
  et~al.}{2018b}]{WaltonS}
{Walton} D.~J.,  et~al., 2018b, \mn@doi [\apjl] {10.3847/2041-8213/aabadc},
  \href {http://adsabs.harvard.edu/abs/2018ApJ...857L...3W} {857, L3}

\bibitem[\protect\citeauthoryear{{Weisskopf}, {Brinkman}, {Canizares},
  {Garmire}, {Murray}  \& {Van Speybroeck}}{{Weisskopf} et~al.}{2002}]{CHANDRA}
{Weisskopf} M.~C.,  {Brinkman} B.,  {Canizares} C.,  {Garmire} G.,  {Murray}
  S.,   {Van Speybroeck} L.~P.,  2002, \mn@doi [\pasp] {10.1086/338108}, \href
  {http://adsabs.harvard.edu/abs/2002PASP..114....1W} {114, 1}

\bibitem[\protect\citeauthoryear{{Wiktorowicz}, {Sobolewska}, {Lasota}  \&
  {Belczynski}}{{Wiktorowicz} et~al.}{2017}]{Wiktorowicz17}
{Wiktorowicz} G.,  {Sobolewska} M.,  {Lasota} J.-P.,   {Belczynski} K.,  2017,
  \mn@doi [\apj] {10.3847/1538-4357/aa821d}, \href
  {http://adsabs.harvard.edu/abs/2017ApJ...846...17W} {846, 17}

\bibitem[\protect\citeauthoryear{{Winter}, {Mushotzky}  \& {Reynolds}}{{Winter}
  et~al.}{2006}]{Winter2006}
{Winter} L.~M.,  {Mushotzky} R.~F.,   {Reynolds} C.~S.,  2006, \mn@doi [\apj]
  {10.1086/506579}, \href
  {https://ui.adsabs.harvard.edu/abs/2006ApJ...649..730W} {649, 730}

\bibitem[\protect\citeauthoryear{{Zombeck}, {Chappell}, {Kenter}, {Moore},
  {Murray}, {Fraser}  \& {Serio}}{{Zombeck} et~al.}{1995}]{CHANDRA_HRC}
{Zombeck} M.~V.,  {Chappell} J.~H.,  {Kenter} A.~T.,  {Moore} R.~W.,  {Murray}
  S.~S.,  {Fraser} G.~W.,   {Serio} S.,  1995, in {Siegmund} O.~H.,  {Vallerga}
  J.~V.,  eds,  \procspie Vol. 2518, EUV, X-Ray, and Gamma-Ray Instrumentation
  for Astronomy VI. pp 96--106, \mn@doi{10.1117/12.218408}

\bibitem[\protect\citeauthoryear{{de Vaucouleurs}, {de Vaucouleurs}, {Corwin},
  {Buta}, {Paturel}  \& {Fouqu{\'e}}}{{de Vaucouleurs}
  et~al.}{1991}]{Vaucouleurs1991}
{de Vaucouleurs} G.,  {de Vaucouleurs} A.,  {Corwin} Jr. H.~G.,  {Buta} R.~J.,
  {Paturel} G.,   {Fouqu{\'e}} P.,  1991, {Third Reference Catalogue of Bright
  Galaxies. Volume I: Explanations and references. Volume II: Data for galaxies
  between 0$^{h}$ and 12$^{h}$. Volume III: Data for galaxies between 12$^{h}$
  and 24$^{h}$.}

\makeatother
\end{thebibliography}

\label{lastpage}
\end{document}